\documentclass[12pt,usenames,dvipsnames,a4paper]{article}

\usepackage{soul}

\usepackage[utf8]{inputenc}
\usepackage[T1]{fontenc}

\usepackage{changepage}
\usepackage{amsmath,amsfonts,amssymb}
\usepackage{epsf,amsmath,bbold,amsfonts,stmaryrd}
\usepackage{mathrsfs}
\usepackage{appendix}
\usepackage{caption}
\usepackage{color}
\usepackage{datetime}
\usepackage{float}
\usepackage{graphicx}
\usepackage[colorlinks]{hyperref}
\hypersetup{pageanchor=false,citecolor=red,urlcolor=red}
\usepackage{indentfirst}
\usepackage[numbers,square,comma,sort&compress,merge]{natbib} 
\usepackage{subfig}
\numberwithin{equation}{section}
\usepackage{mathtools}
\usepackage[colorinlistoftodos]{todonotes}
\usepackage{ytableau}
\usepackage{tabu}
\usepackage{esvect}

\usepackage{calc}

\allowdisplaybreaks

\def\a{\alpha}

\def\c{\chi}

\def\d{\delta}
\def\D{\Delta}

\def\eps{\varepsilon}
\def\f{\frac}
\def\g{\gamma}

\def\G{\Gamma}

\def\l{\left}

\def\m{\mu}
\def\n{\nu}

\def\nn{\nonumber}

\def\p{\partial}
\def\vp{\varphi}

\def\r{\right}
\def\s{\sigma}
\def\t{\theta}

\def\vp{\varphi}

\def\x{\xi}

\def\z{\zeta}

\def\be{\begin{equation}}
\def\ee{\end{equation}}

\def\bea{\begin{eqnarray}}
\def\eea{\end{eqnarray}}

\def\ba{\begin{array}}
\def\ea{\end{array}}

\def\bc{\begin{center}}
\def\ec{\end{center}}

\def\bl{\begin{flushleft}}
\def\el{\end{flushleft}}

\def\br{\begin{flushright}}
\def\er{\end{flushright}}

\def\bi{\begin{itemize}}
\def\ei{\end{itemize}}

\def\bt{\begin{tabular}}
\def\et{\end{tabular}}

\makeatletter

\newsavebox\myboxA
\newsavebox\myboxB
\newlength\mylenA

\newcommand*\xoverline[2][0.75]{%
    \sbox{\myboxA}{$\m@th#2$}%
    \setbox\myboxB\null
    \ht\myboxB=\ht\myboxA%
    \dp\myboxB=\dp\myboxA%
    \wd\myboxB=#1\wd\myboxA
    \sbox\myboxB{$\m@th\overline{\copy\myboxB}$}
    \setlength\mylenA{\the\wd\myboxA}
    \addtolength\mylenA{-\the\wd\myboxB}%
    \ifdim\wd\myboxB<\wd\myboxA%
       \rlap{\hskip 0.5\mylenA\usebox\myboxB}{\usebox\myboxA}%
    \else
        \hskip -0.5\mylenA\rlap{\usebox\myboxA}{\hskip 0.5\mylenA\usebox\myboxB}%
    \fi}
\makeatother

\def\be{\begin{equation}}
\def\ee{\end{equation}}

\def\bea{\begin{eqnarray}}
\def\eea{\end{eqnarray}}

\def\f{\frac}

\def\p{\partial}

\usepackage{xspace}
\newcommand*{\ie}{i.e., }
\newcommand*{\eg}{e.g., }
\newcommand*{\fig}{fig.\@\xspace}
\newcommand*{\eq}{eq.\@\xspace}
\newcommand*{\eqs}{eqs.\@\xspace}
\newcommand*{\lhs}{l.h.s.\@\xspace}
\newcommand*{\rhs}{r.h.s.\@\xspace}
\newcommand*{\cf}{cf.\@\xspace}
\newcommand*{\wlg}{w.l.o.g.\@\xspace}

\usepackage{xifthen}
\newcommand*\diff{\mathrm{d}} 
\newcommand*\ldiff[2][]{ \ifthenelse{\isempty{#1}}{ \diff #2}{\diff^#1#2} \,} 
\let\limitint\int 
\renewcommand{\int}{\limitint \!} 

\begin{document}

\begin{titlepage}

\vspace*{-2.5cm}
\begin{adjustwidth}{}{-.45cm}
\br
{
\begin{tabular}{@{}l@{}}
\small LMU--ASC~25/23
 \end{tabular}
 }
\er
\end{adjustwidth}

\vspace*{1.5cm}
\begin{adjustwidth}{-1.3cm}{-.7cm}

\begin{center}
    \bf \Large{Scale invariant Einstein-Cartan gravity\\ 
and flat space conformal symmetry
        }
\end{center}
\end{adjustwidth}

\begin{center}
\textsc{Georgios K. Karananas,$^\star$~Mikhail Shaposhnikov,$^\dagger$~Sebastian Zell\,$^\ddagger$}
\end{center}

\begin{center}
\it {$^\star$Arnold Sommerfeld Center\\
Ludwig-Maximilians-Universit\"at M\"unchen\\
Theresienstra{\ss}e 37, 80333 M\"unchen, Germany\\
\vspace{.4cm}
$^\dagger$Institute of Physics \\
\'Ecole Polytechnique F\'ed\'erale de Lausanne (EPFL) \\ 
CH-1015 Lausanne, Switzerland\\
\vspace{.4cm}
$^\ddagger$Centre for Cosmology, Particle Physics and Phenomenology -- CP3,\\
	Universit\'e catholique de Louvain,\\
 B-1348 Louvain-la-Neuve, Belgium
}
\end{center}

\begin{center}
\small
\texttt{\small georgios.karananas@physik.uni-muenchen.de}  \\
\texttt{\small mikhail.shaposhnikov@epfl.ch} \\
\texttt{\small sebastian.zell@uclouvain.be} 
\end{center}

\begin{abstract}
We find the conditions under which scale-invariant Einstein-Cartan gravity with scalar matter fields leads to an approximate conformal invariance of the flat space particle theory up to energies of the order of the Planck mass. In the minimal setup, these models, in addition to the fields of the Standard Model and the graviton, contain only one extra particle -- a massless dilaton. Theories of this type can pave the way for a self-completion all the way up the Planck scale and lead to rather universal inflationary predictions, close to those of the simplest Higgs-inflation scenario in the metric theory of gravity.

\end{abstract}

\end{titlepage}

\section{Introduction}

In~\cite{Karananas:2021gco} we constructed a class of phenomenologically viable theories based on the  Einstein-Cartan (EC) gravity \cite{Cartan:1922,Cartan:1923,Cartan:1924,Cartan:1925,Einstein:1925, Einstein:1928,Einstein:19282} which enjoy exact but spontaneously broken scale invariance. These theories, in addition to the Standard Model (SM)  and graviton fields, contain just one additional particle -- a massless dilaton, being the Nambu-Goldstone (NG) boson of this symmetry.  The classical action of these theories is selected with the use of systematic requirements that aim at capturing the minimal ambiguity inevitably contained in General Relativity (GR) \cite{Karananas:2021zkl}. This is achieved by demanding equivalence to the metric formulation of GR in the absence of matter while at the same time avoiding assumptions as far as possible (see \cite{Karananas:2021zkl,Karananas:2021gco,Rigouzzo:2022yan,Rigouzzo:2023sbb} for detailed discussions). The criteria of \cite{Karananas:2021zkl} can be expressed as follows \cite{Rigouzzo:2023sbb}:\footnote{The requirements of \cite{Karananas:2021zkl} were refined in \cite{Rigouzzo:2023sbb} but both sets of conditions are equivalent for the theories considered in the present paper.}
\begin{itemize}
 \item[\emph{i)}]   The action is polynomial with respect to all matter fields and  curvature invariants.
  \item[\emph{ii)}] The action must not contain operators with more than two derivatives (where torsion counts as one derivative).
 \item[\emph{iii)}] The theory should be scale-invariant and thus only contain dimensionless parameters. 
\end{itemize}
In the algebraic flat spacetime limit defined as $e_\m ^A  =  \d_\m^A,~\omega_\m^{AB} = 0$, conditions \emph{i)} - \emph{iii)} imply that the theory enjoys not only scale-invariance but a  wider symmetry -- invariance under the 15-parameter full  conformal group.\footnote{Here $e_\m^A$ is the tetrad/vierbein and $\omega_\m^{AB}$ is the (spin) connection. As usual, Greek letters are employed for spacetime indices, while capital Latin letters for Lorentz indices.}

The {\em physical} low-energy limit (we stress that this is not the same as the algebraic flat spacetime limit defined above) of these theories is derived by integrating out the non-dynamical torsion and going to the Einstein frame, such that the (canonical) graviton is disentangled from the scalar degrees of freedom. The resulting  physical particle physics action is found by dropping the Einstein-Hilbert term and making the metric flat; in general, it is a non-polynomial function of the scalar fields. This action captures the dynamics of a {\em scale invariant}  theory, where this symmetry is broken spontaneously (realized nonlinearly). For different aspects of this construction, such as quantum corrections, its cosmology and phenomenology, see e.g.~\cite{Wetterich:1987fk,Wetterich:1987fm,Dehnen:1992jc,Wetterich:1994bg,Cervantes-Cota:1995ehs,Foot:2007iy,Shaposhnikov:2008xb,Shaposhnikov:2008xi,Shaposhnikov:2008ar,Shaposhnikov:2009nk,GarciaBellido:2011de,Blas:2011ac,GarciaBellido:2012zu,Bezrukov:2012hx,Monin:2013gea,Tavares:2013dga,Khoze:2013uia,Csaki:2014bua,Rubio:2014wta,Ghilencea:2015mza,Karam:2015jta,Trashorras:2016azl, Karananas:2016grc,Ferreira:2016vsc,Karananas:2016kyt,Karam:2016rsz,Ferreira:2016wem,Ferreira:2016kxi,Ghilencea:2016dsl,Shkerin:2016ssc,Rubio:2017gty,Tokareva:2017nng,Casas:2017wjh,Ferreira:2018itt,Ferreira:2018qss,Shaposhnikov:2018jag,Burrage:2018dvt,Lalak:2018bow,Gorbunov:2018llf,Iosifidis:2018zwo,Casas:2018fum,Shkerin:2019mmu,Herrero-Valea:2019hde,Karananas:2019fox,Rubio:2020zht,Karananas:2020qkp,Hill:2020oaj,Piani:2022gon,Karananas:2021gco}~for a far-from-complete list of relevant works.  

The fact that the resulting particle physics theory for arbitrary choice of different non-minimal couplings to gravity  is only scale, but not conformally invariant means that the interaction of the scalar degree of freedom hidden in the metric (or in the vierbein and connection if we talk about EC gravity) breaks explicitly the conformal invariance.  

The aim of the present work is to single out the subclass of the theories defined by~\emph{i)} - \emph{iii)} by adding an extra requirement: {\em the resulting physical theory for energies  up to the Planck scale should approximately be conformally invariant rather than only being scale invariant}. We note here that it does not make much sense to require the existence of  exact conformal symmetry for all energies, since irrespectively of nonminimal coupling(s), the mere existence of gravitational interactions is in conflict~\cite{Shaposhnikov:2022dou,Shaposhnikov:2022zhj} with having this symmetry in the physical low-energy limit due to the Weyl anomaly (for a review see~\cite{Duff:1993wm}). With the use of the language of effective field theories, the physical action of the theory constructed along the lines above is required to read as
\be
S_{\rm eff}= S_{\rm conf} + \sum_{n=2}^\infty \frac{1}{\Lambda_{\rm conf}^n} S_{\rm scale}^{(n)} \ .
\label{decomp}
\ee 
The first term $S_{\rm conf}$ is essentially a non-polynomial conformally invariant action with the symmetry nonlinearly realized, while the breaking pieces  $S_{\rm scale}^{(n)}$ are suppressed by powers of the scale $\Lambda_{\rm conf}$. For reasons that will become clear shortly, we shall take $\Lambda_{\rm conf}= M_P$ (it is assumed that the dimensionless coefficients appearing in $S_{\rm scale}^{(n)}$ are of the order of one).

\begin{figure}[!t]
    \centering
    \includegraphics[scale=.4]{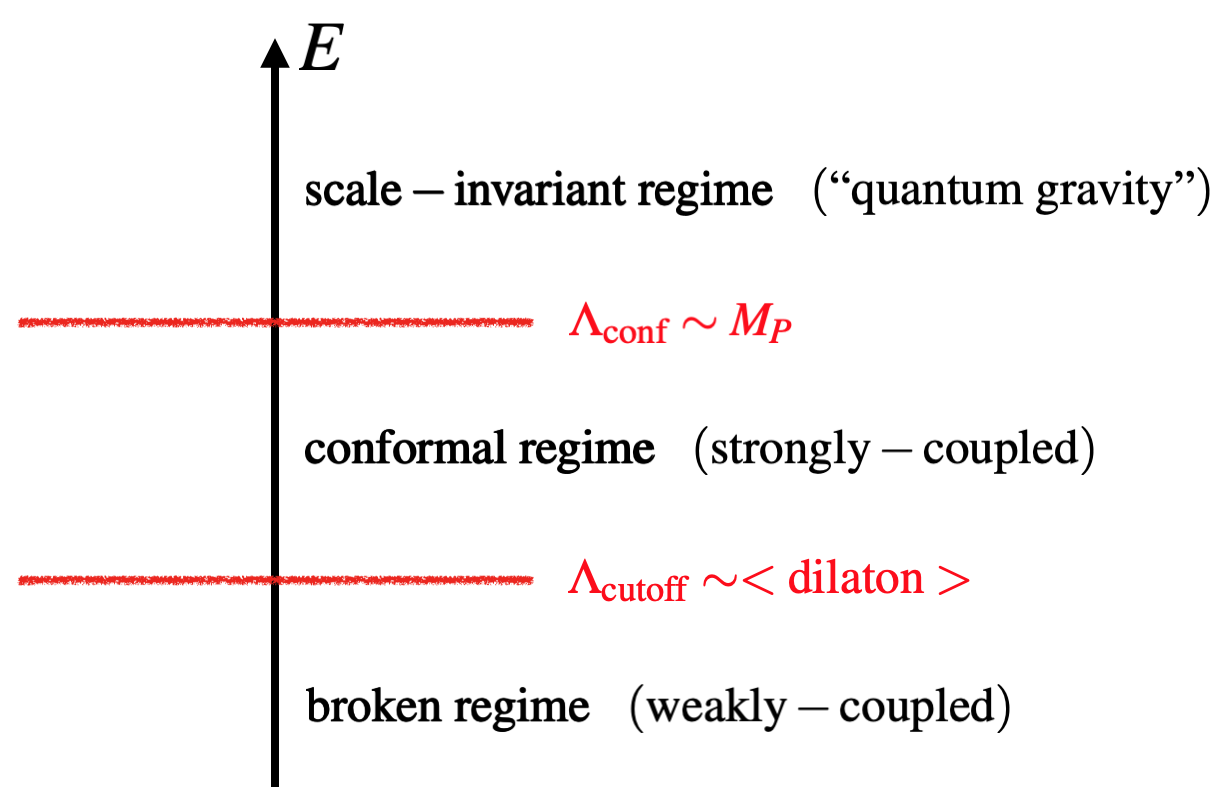}
    \caption{The phase portrait of the theory.}
    \label{fig:phase}
\end{figure}

 Our interest in this class of theories stems from various field-theoretical as well as phenomenological considerations, on which we elaborate now.

 First,  theories with exact conformal invariance (CFT's) exhibit ultraviolet (UV) fixed points in the renormalization group running and are UV complete (for a review see~\cite{Rychkov:2016iqz}).  As already mentioned, the action $S_{\rm conf}$, to be constructed in what follows, is non-polynomial and thus non-renormalizable. As such, it has an intrinsic dimensionful parameter $\Lambda_{\rm cutoff}$ (related to the dilaton vev)  which can in principle be much smaller than the Planck scale.  If there were no ``hidden'' conformal symmetry, one would expect that the theory defined by $S_{\rm conf}$ requires some sort of UV completion around the vicinity of $\Lambda_{\rm cutoff}$ with new degrees of freedom showing up at these energies. It is plausible that the conformal symmetry improves the situation, \ie the theory may ``self-complete'' in the energy interval $\Lambda_{\rm cutoff} < E \lesssim  \Lambda_{\rm conf}$, entering a strongly-coupled conformal regime described by an unbroken CFT.

Second, there are arguments that the requirement of unitarity applied to quantum field theory in flat spacetime excludes theories with scale but no conformal invariance~\cite{Luty:2012ww}. Taken at face value, this would mean that if the energy scale $\Lambda_{\rm conf}$ around which conformal symmetry is explicitly broken down to scale symmetry is much smaller than $M_P$, $\Lambda_{\rm conf}\ll M_P$,  then the  physical action~(\ref{decomp}) as well as the initial action defined by the requirements~\emph{i)} - \emph{iii)} would not make sense for energies exceeding $\Lambda_{\rm conf}$. Even the part of the theory without gravity should be modified in the energy interval $\Lambda_{\rm conf} < E < M_P$, probably by integrating-in new physics. If on the contrary $\Lambda_{\rm conf} \simeq M_P$, then the theory merges (in some way) with gravity respecting only the scale-symmetry. The region of validity of (\ref{decomp}) extends all the way to the Planck scale, and thus no new degrees of freedom are required until such energies. Arguably, this is a rather non-trivial expectation.  It remains to be seen if it is true.

For clarity let us recapitulate how we envisage the behavior of the theories under consideration from a bottom-up perspective; see also \fig~\ref{fig:phase} for a graphical account. At energies below the dilaton expectation value, i.e. for $E\lesssim\Lambda_{\rm cutoff}$, the theory is in its weekly-coupled regime, with conformal symmetry spontaneously broken and the accompanying massless dilaton present. For higher energies, $\Lambda_{\rm cutoff}<E\lesssim \Lambda_{\rm conf}\sim M_P$,  the  theory is in a strongly-coupled regime with restored conformal symmetry and thus becomes a full-blown CFT. Finally, gravity kicks in at $E\sim M_P$, explicitly violating conformality but preserving scale-invariance.   

In addition to the field-theoretical arguments presented so far, the third and final motivation for investigating aspects of such theories comes from cosmological phenomenology. Without the requirement of approximate conformal invariance up to the Planck scale, theories constructed according to the criteria \emph{i)} to  \emph{iii)}(or even without the requirement of scale-invariance) contain numerous a priori unknown coupling constants and generically it becomes impossible to derive unique observable predictions \cite{Langvik:2020nrs, Shaposhnikov:2020frq, Shaposhnikov:2020gts,Karananas:2021zkl,Karananas:2021gco}. We find indications that approximate conformal symmetry improves the situation considerably: in the context of inflationary model building, we therefore conjecture that primordial observables become nearly universal and close to the ones of single-field Higgs inflation in the metric formulation of General Relativity \cite{Bezrukov:2007ep}. As an additional bonus, this leads to excellent agreement with observations \cite{Planck:2018jri,BICEP:2021xfz}. As a proof of concept we provide a concrete example and we leave a detailed investigation for future work.

This article is organized as follows. In Sec.~\ref{sec:single_scalar}, we review briefly well-known facts associated with the constraints conformal symmetry imposes on the action of a real scalar field. In Sec.~\ref{sec:multifield}, we generalize these findings to the case of $N$ real scalar fields. The fact that conformal symmetry is more restrictive than scale symmetry becomes apparent already at the level of two derivatives: the kinetic sector is fixed in a rather nontrivial manner---this is to be contrasted with the single field case, where to see such differences one needs to go to the fourth order in the derivatives of the field. We also show that nonlinearly realized conformal symmetry, that is when scales are generated, also constrains the $N$-field action nontrivially.  In Sec.~\ref{sec:gravity}, we discuss how the explicit breaking of conformal symmetry that goes hand-in-hand with the inclusion of (dynamical) gravity manifests itself at the level of the action. In Sec.~\ref{sec:higgs_inflation}, we conjecture that flat-space conformality (together with requiring agreement with observations) is powerful enough to make the various Higgs-dilaton models almost indistinguishable from single-field metric Higgs inflation scenario as far as inflationary predictions are concerned.  In Sec.~\ref{sec:conclusions}, we summarize our findings and conclude. Appendix~\ref{app:deviation} discusses how the inflationary dynamics are altered once the selection criteria~\emph{i)} - \emph{iii)} are relaxed. 

\section{An invitation: a single real scalar field}
\label{sec:single_scalar}

In this section, we shall discuss well-known facts (see e.g.~\cite{Jackiw:2011vz,Hinterbichler:2022ids}) about how linearly realized invariance under the full conformal group completely fixes the action of a real scalar field. We do that in two different approaches: algebraic and geometric.  In the former, one writes down the most general action involving a scalar and then investigates how it is constrained by requiring that it be invariant under the conformal variation of the field. In the latter, one ``dresses'' the Minkowski metric with a scalar field and then takes the action to be a derivative expansion of diffeomorphism-invariant objects constructed out of this conformally flat metric with arbitrary coefficients.  The following discussion is straightforward and contains nothing new, but we include it for the sake of making the article as self-contained as possible.

\subsection{Algebraic approach}
\label{sec:algebraic_approach}

Consider a real scalar field $\varphi$ with scaling dimension $\D_\vp\neq 0$ in a $D>2$ dimensional\,\footnote{It is well known that conformal symmetry for $D=2$ has a number of peculiarities which make it a separate topic of investigations.} Minkowski spacetime with metric $\eta_{\m\n}={\rm diag}(-1,1,\ldots,1)$. We take the action to contain terms that are at most quadratic in the derivatives of the field,
\be
\label{eq:actionSingleField}
S = -  \int \diff^Dx~\l[\f 1 2 G(\varphi)(\p_\m\varphi)^2 +V(\varphi) \r] \ ,
\ee
with $G(\varphi)$ and $V(\varphi)$ arbitrary functions. Note that noncanonical kinetic terms appear naturally whenever gravity enters the picture, being the aftermath of nonminimal coupling(s).

 Since $\varphi$ is the only dimensionful quantity (apart from $\p_\m$), it follows from requiring invariance under dilatations only---equivalently by performing dimensional analysis---that
\be 
\label{eq:conformalPotentialSingleField}
V(\vp) = \tilde c \vp^\f{D}{\D_\vp} \ ,~~~\tilde c = {\rm constant} \ ,
\ee
as well as
\be
\label{eq:conformalKineticSingleField}
G(\vp) = c \vp^\a \ ,~~~c ={\rm constant} \ ,
\ee
where
\be
\label{eq:alpha}
\a = -\f{2}{\D_\varphi}\l(\D_\varphi-\f D 2 +1 \r) \ .
\ee
Plugging these results into $S$, we end up with  
\be 
\label{eq:conformalActionSingleField}
S= -\int \diff^Dx\l[\f 1 2 c\vp^\a (\p_\m \vp)^2 +\tilde c \vp^\f{D}{\D_\vp} \r] \ .
\ee

Next, we study the constraints imposed upon the action~\eqref{eq:actionSingleField} by requiring that it be conformally invariant, i.e. for $S$ to have a vanishing  conformal variation
\be
\d_c S = \int \diff^Dx\,\f{\d S}{\d\varphi}\d_c\varphi = 0 \ ,
\ee
with 
\begin{equation}
\delta_c \varphi = - \left( \epsilon ^ \mu \partial _ \mu \varphi  
  +\frac{\Delta_{\varphi} }{D} \varphi  \partial_ \mu \epsilon ^\mu \right) \ .
\end{equation}
Here $\epsilon ^ \mu $ is the conformal Killing vector (more details may for instance be found in~\cite{DiFrancesco:1997nk,Rychkov:2016iqz})
\begin{equation}
  \label{eq:conf_killing}
\epsilon ^ \mu = a ^ \mu + \omega ^ \mu _ {\ \nu} x ^ \nu + c x ^ \mu
+ 2 (b\cdot x) x ^ \mu - x ^ 2 b ^ \mu \ ,
\end{equation}
where $a ^ \mu,\omega _ {\mu\nu}= -\omega_{\nu\mu},c$ and $b ^ \mu$
are constant parameters associated with translations, Lorentz
transformations,  dilatations and special conformal transformations, respectively; Eq.~(\ref{eq:conf_killing}) is 
the solution to the flat spacetime conformal Killing equation 
\be
\label{eq:conf_kill_eq}
\p_\m \epsilon_\n + \p_\n \epsilon_\m = \f 2 D \eta_{\m\n}\p_\lambda \epsilon^\lambda \ .
\ee

A straightforward computation shows that 
\bea
\d_c S &=& -\int\diff^Dx  \Bigg[\f 1 2 G'(\varphi)\d_c\varphi(\p_\m\varphi)^2+G(\varphi) \p_\m\varphi \p^\m\d_c\varphi +V'(\vp)\d_c\vp \Bigg]  \nn\\
&=&\int\diff^Dx \Bigg\{ \f {\p_\n\epsilon^\n}{D} \Bigg[\l( \l(\D_\varphi-\f D 2 + 1  \r)G(\varphi) +\f{\D_{\varphi}}{2}\varphi G'(\varphi)\r)\p_\m \varphi \p^\m \varphi \nn\\
&+&D V(\vp) -\D_\vp \vp V'(\vp)\Bigg] -\f{\D_\varphi}{2 D}\p^\m\p_\n\epsilon^\n\varphi^2G'(\varphi)\p_\m\varphi \Bigg\} \ , 
\label{eq:conf_var}
\eea
where we arranged the terms with increasing powers of derivatives acting on the conformal Killing vector $\epsilon_\m$.
For the potential, we notice that it is fixed to be a homogeneous function of the field 
\be
V(\vp) = \f{\D_\vp}{D}\vp V'(\vp) \ .
\ee
which is automatically fulfilled due to dimensional analysis, as follows from \eq \eqref{eq:conformalPotentialSingleField}. We make an analogous observation with regard to the non-canonical coefficient of the kinetic term. Because of \eq \eqref{eq:conformalKineticSingleField}, the last term in~(\ref{eq:conf_var}) becomes a total derivative, which can be immediately verified by using the fact that $\epsilon$ is at most quadratic in $x$. Then  it follows from \eq \eqref{eq:alpha} that the remaining contributions involving $G(\varphi)$ vanish in \eq \eqref{eq:conf_var}. This shows that in the case of a single field, conformal invariance follows automatically from scale invariance.

Finally, one may make the kinetic term of the field in the action~\eqref{eq:conformalActionSingleField} canonical by introducing 
\be
\label{eq:field_canon}
\bar{\varphi}= 
\f{2\sqrt {c}\D_\vp }{D-2}\vp^\f{D-2}{2\D_\vp}   \ ,
\ee
such that 
\be
\bar S = -\int \diff^Dx \l[ \f{1}{2} (\p_\m\bar\vp)^2+\f{D-2}{2D}\lambda \bar\vp^\f{2D}{D-2} \r] \ ,
\ee
with 
\be
\lambda = \f{2D}{D-2}\tilde c \l(\f{D-2}{2\sqrt{c}\D_\vp}\r)^\f{2D}{D-2} \ . 
\ee
We remark that a canonical kinetic term immediately translates into $\bar\vp$ having canonical scaling dimension too, i.e.
\be
\D_{\bar \vp} = \f D 2 -1 \ ,
\ee
as it may be verified from its definition eq.~(\ref{eq:field_canon}).   

\subsection{Geometric approach}
\label{sec:geometric_approach}

We can equivalently obtain the same action by utilizing the standard trick of first identifying the field with the scalar mode of a conformally flat metric and then writing down the invariants at each order in derivatives with arbitrary coefficients, see for instance~\cite{Nicolis:2008in,Hinterbichler:2012mv,Goon:2012dy,Hinterbichler:2022ids}.

Since our purpose here is to only get terms at most quadratic in derivatives of the field, our action comprises two terms and reads 
\be
\label{eq:grav_act_single_field}
S = - \int \diff^Dx \sqrt{H} \l( \f{c \D_\vp^2}{2(D-1)(D-2)}  M^{D-2} {\mathcal R[H]} +\tilde c M^D   \r ) \ , 
\ee
where as before $\tilde c, c$ are (dimensionless) constants, $M$ is a mass parameter, while $H=(-1)^{D+1}{\rm det}(H_{\m\n})$ and $\mathcal R[H]$ are the  Ricci scalar of the dressed metric 
\be
H_{\m\n} = \omega^2 \eta_{\m\n} \ ,~~~\omega = M^{-1}\vp^\f{1}{\D_\vp} \ . 
\ee

Note that since $\mathcal R[H]$ is the kinetic term of the conformal mode of the metric in disguise, its sign has to  be chosen opposite of what it would be had gravity been dynamical. Indeed,~using the standard expressions (see for example Ref.~\cite{Carroll:2004st})
\be
\label{eq:dressed_metric}
H = \omega^{2D} \ ,~~~\mathcal R[H] = -(D-1)\l(2\omega^{-3} \p^2\omega +(D-4)\omega^{-4} (\p_\m\omega)^2\r) \ ,  
\ee
we find after a straightforward computation that the action boils down to 
\be
S = -\int \diff^Dx \l (\f 1 2 c\vp^\a (\p_\m\vp)^2 +  \tilde c \vp^\f{D}{\D_\vp}\r) \ ,
\ee
which is of course identical to what we got with the algebraic method, see~(\ref{eq:conformalActionSingleField}). 
 Had the sign of the scalar curvature been ``plus,'' the kinetic term for the dilaton would correspond to a ghost. 

\section{Multifield generalization}
\label{sec:multifield}

We now turn to the multifield generalization of the findings in the previous section. We start from a linear realization of conformal symmetry and construct the most general conformally invariant action comprising $N$ real scalar fields in a $D>2$ flat spacetime,  giving up condition~\emph{i)}~formulated in the Introduction. To the best of our knowledge, such Lagrangians, i.e. with more than one scalar field, have not been presented/constructed before, at least not in their full generality. Our strategy is to write down the most general action that is at most quadratic in the derivatives of the various fields and then require that this be invariant under dilatations as well as special conformal transformations. As far as terms not involving derivatives are concerned, dilatations are enough to completely fix the potential to be a homogeneous function of the fields. For the kinetic sector of the theory, we find that conformal invariance puts more severe restrictions than scale invariance; this is in contradistinction with what happens with a single field, where at the level of two derivatives requiring invariance under special conformal transformations does not bring any new information. More specifically, we observe that certain coefficient functions are not independent but are interrelated to each other, the aftermath of imposing invariance under special conformal transformations. For the most minimal situation in which the number of fields is $N=2$, a conformally invariant kinetic sector is in a one-to-one with a vanishing curvature for the manifold spanned by the derivatives of the fields, i.e. a flat target manifold. One may easily convince oneself that this is not the case for theories invariant under dilatations only. Therefore, a useful criterion to distinguish between these different situations is to compute the curvature of the two-dimensional kinetic sector.  Then we repeat this program for a nonlinearly realized conformal symmetry, to prepare the ground for the inclusion of gravity.

\subsection{Linear realization}

 Let us consider $N$ real scalar fields $\t_1,\ldots,\t_N$. To simplify the following computations, we employ \wlg an ``angular parametrization'' by singling out one of the fields, say $\t_1=\varphi$, in that its scaling dimension will be $\D_\varphi\neq 0$, while the rest $N-1$ fields $\t_i,~i=2,\ldots,N$ have $\D_i = 0$. Then the most general action that includes terms with at most two derivatives of the fields reads
 \be
 \label{eq:actionMultiField}
S =  -\int \diff^Dx~\Big( K  +V \Big)\ ,
\ee
where $V=V(\phi_I)$ stands for the potential, while 
\be
K = \f 1 2 G_{IJ} \p_\m \phi_I \p^\m \phi_I \ ,
\ee
is the kinetic sector of the theory; to keep the expression compact we introduced $\phi_I=(\varphi,\t_2,\ldots,\t_N)$ and $G_{IJ}=G_{IJ}(\phi_K)$ is a real, symmetric, non-singular $N\times N$ matrix---the metric of the target  manifold; explicitly, 
\be 
\label{eq:metric_explicit}
G_{IJ} = \begin{pmatrix}
    G_{\vp\vp} & G_{\vp i} \\
    G_{\vp i} & G_{ij} 
\end{pmatrix} \ .
\ee
As usual, summation over all repeated indexes is tacitly assumed.

\subsubsection{Algebraic approach}
As before, we will at first only require invariance under dilatations. Since $\varphi$ is the only field with non-vanishing mass dimension, it follows that
\begin{equation}
\label{eq:conformalPotential}
    V(\varphi;\t_j) = \varphi^{\f{D}{\D_\varphi}}v(\t_j) \ ,
\end{equation}
as well as 
\be
\label{eq:coef_dilatations_only}
G_{\varphi\varphi} = \varphi^\a F_{\varphi\varphi} (\t_i) \ ,~~G_{\varphi i} =   \varphi^{\a+1} F_{\varphi i} (\t_j) \ ,~~G_{ij} = \varphi^{\a+2} F_{ij} (\t_k) \ ,
\ee
with $\a$ defined previously, cf.~(\ref{eq:alpha}), while  $v$ and $F$'s are arbitrary functions of $\t$.

Next, to understand whether conformal symmetry imposes extra constraints we proceed with the full variation of the action by utilizing 
 \begin{equation}
  \label{eq:conf_trans}
\delta_c \varphi = - \left( \epsilon ^ \mu \partial _ \mu \varphi  
  +\frac{\Delta_{\varphi} }{D} \varphi  \partial_ \mu \epsilon ^\mu \right) \ ,~~~\delta_c \t_i =-   \epsilon ^ \mu \partial _ \mu \t_i \ .
\end{equation}

Let us concentrate on the non-derivative terms comprising the potential of the theory.  Under~(\ref{eq:conf_trans}), we observe that 
\begin{align}
\delta_c \int \diff^Dx~  V(\varphi;\t_j) 
  & = \int \diff^Dx~\partial _ \mu \epsilon ^ \mu \left( V (\varphi;\t_j) -\frac{\D_\varphi}{D} \varphi   \frac{\partial V(\varphi;\t_j)}{\partial
                   \varphi }\right)  \ ,
\end{align}
after integrating by parts  and dropping a total derivative. Since $\partial_\mu \epsilon^\mu \neq 0$ as follows
from~(\ref{eq:conf_killing}), for the conformal variation of the
potential to vanish we have to require that 
\begin{equation}
\label{eq:potential_intermediate}
V (\varphi;\t_j) =\frac{\D_\varphi}{D} \varphi   \frac{\partial V(\varphi;\t_j)}{\partial
                   \varphi }\ ,
\end{equation}
with $v(\t_j)$ an arbitrary function of the argument. Evidently, this condition is fulfilled by \eq \eqref{eq:conformalPotential}.

We turn to the kinetic sector of the multifield theory, 
\be
\label{eq:conf_K_start0}
\int \diff^Dx~K =\f 1 2\int \diff^Dx~ G_{IJ} \p_\m \phi_I \p^\m \phi_I \ ,
\ee
which upon using~(\ref{eq:metric_explicit}), can be expanded as
\be
\label{eq:conf_K_start}
\int \diff^Dx~K = \f 1 2\int \diff^Dx~\left[ G_{\varphi\varphi}\p_\m\varphi \p^\m \varphi +2G_{\varphi i}\p_\m\varphi \p^\m \t_i +  G_{ij}\p_\m \t_i\p^\m \t_j\right] \ .
\ee
Then,
\bea
\d_c \int \diff^Dx~K &=& -
\int\diff^Dx\Bigg\{ \f {\p_\n\epsilon^\n}{D} \Bigg[\l(\l(\D_\varphi-\f D 2 + 1  \r)G_{\varphi\varphi} +\f{\D_{\varphi}}{2}\varphi\f{\p G_{\varphi\varphi}}{\p \varphi}\r)\p_\m \varphi \p^\m \varphi \nn \\
&+&\l( \l(\D_\varphi-D + 2  \r)G_{\varphi i} + \D_{\varphi}\varphi\f{\p G_{\varphi i}}{\p \varphi}\r)\p_\m \varphi \p^\m \t_i\nn\\
&+& \l(\l(-\f{D}{2}+1\r)G_{ij}+\f {\D_\varphi}{2} \varphi \f{\p G_{ij}}{\p \varphi}\r)\p_\m\t_i\p^\m \t_j \Bigg] \nn\\
&-&\f{\D_\varphi}{2 D}\p^\m \p_\n\epsilon^\n\Bigg(\varphi^2\ \f{\p G_{\varphi\varphi}}{\p \varphi}\p_\m \varphi+\varphi\l(\varphi\f{\p G_{\varphi\varphi}}{\p \t_i} -2 G_{\varphi i}\r)\p_\m \t_i\Bigg)\Bigg\} \ . 
\label{eq:conf_kinetic}
\eea
As before, we have arranged the various terms in increasing derivatives of $\p_\m \epsilon^\m$. We observe that the first three lines vanish for the $G_{IJ}$'s defined as in \eq \eqref{eq:coef_dilatations_only}. Plugging this into the last line of the conformal variation of $K$~(\ref{eq:conf_kinetic}) and setting it to zero, we get 
\be
\label{eq:notindependentmixing}
F_{\varphi i}= \f{1}{\a+2}\f{\p F_{\varphi\varphi}}{\p \t_i} \ ,  
\ee
that is, the mixing function is not  independent, but rather related nontrivially to the gradient of $F_{\varphi\varphi}$. 

\vskip.3cm

\emph{We conclude that in the case of multiple fields, conformal invariance leads to additional non-trivial constraints in addition to the ones that follow from dilatations only.}

\vskip.3cm

Plugging what we found in the starting point, see \eq \eqref{eq:actionMultiField}, we end up with
\be
\label{eq:conf_K_intermediate}
\int \diff^Dx~K =  \f 1 2\int \diff^Dx\l[ \varphi^\a F_{\varphi\varphi}\p_\m \varphi \p^\m \varphi +\f{2}{\a+2}\varphi^{\a+1}\f{\p F_{\varphi\varphi}}{\p \t_i}\p_\mu\varphi \p^\m\t_i + \varphi^{\a+2} F_{ij}\p_\m\t_i\p^\m\t_j   \r] \ .
\ee
Exactly like we did in the single field case, we can redefine the $\vp$ field such that its scaling dimension becomes the canonical one---this is achieved in terms of $\bar\vp$ defined in~(\ref{eq:field_canon}) with $c=1$. We find
\be
\int \diff^Dx~K = \f 1 2 \int \diff^Dx \l[F_{\varphi\varphi}(\p_\m \bar \varphi)^2+ \bar\varphi \f{\p F_{\varphi\varphi}}{\p \t_i}\p_\mu\bar\varphi \p^\m\t_i+\l(\f{\a+2}{2}\r)^2\bar\varphi^2 F_{ij}\p_\m\t_i\p^\m\t_j  \r] \ .
\ee
We now go a step further by untangling the kinetic terms of $\bar\varphi$ and $\t_i$'s and actually canonicalizing the former all in a single shot. To this end, it suffices to simply introduce 
\be
\c = \sqrt{F_{\varphi\varphi}}\bar\varphi \ ,
\ee
in terms of which we immediately find
\be
\label{eq:kin_canonical_dilaton}
\int \diff^Dx~K = \f 1 2 \int \diff^Dx\l[ \p_\m \c \p^\m \c +\c^2 f_{ij}\p_\m\t_i\p^\m\t_j   \r] \ , 
\ee
with 
\be
f_{ij}= \f{1}{4F_{\varphi\varphi}}\l( (\a+2)^2F_{ij} - \f{1}{F_{\varphi\varphi}} \f{\p F_{\varphi\varphi}}{\p \t_i}\f{\p F_{\varphi\varphi}}{\p \t_j}\r)  \ .
\ee
This corresponds to the block-diagonal field-space metric\,\footnote{  Interestingly, this exact kinetic structure appeared recently in~\cite{Zwicky:2023fay}, in an attempt to rectify the  
 (non-) improvement of the Nambu-Goldstone modes associated with the breaking of internal symmetries. In that specific context, our $\t_i$ fields of vanishing scaling dimension correspond to pions $\pi_a$ and $f_{ij}\to \d_{ab}$. }
\be \label{eq:field_space_metric_diagonal}
G^{(\chi)}_{IJ}= 
\begin{pmatrix}
    1 & 0 \\
    0 & \c^2 f_{ij} 
\end{pmatrix} \ .
\ee 
At the same time, changing variables from $\bar \varphi$ to $\c$ results into the potential being nontrivially modified as 
\be
\label{eq:pot_canonical_dilaton}
V = \c^\f{2D}{D-2}\tilde v(\t_i) \ ,
\ee
with 
\be
\tilde v(\t_i) = \l(\f{D-2}{2\sqrt{c}\D_\vp}\r)^\f{2D}{D-2}F_{\varphi\varphi}^{-\f{D}{D-2}}(\t_i)v(\t_i)  \ .
\ee

Before moving on, let us discuss what would change had we confined ourselves to invariance under dilatations only. The coefficient functions would then be given by~(\ref{eq:coef_dilatations_only}), but the mixing functions would not be related to the gradients of $F_{\vp\vp}$. Nevertheless, it is still possible to single out one of the fields by block-diagonalizing the kinetic sector in terms of 
\be
\label{eq:dilatations_1}
\tilde \c =  \f{\bar \vp}{\Phi(\t_i)} \ ,
\ee
where
\be
\f{\p\Phi(\t_i)}{\p\t_j} = -\f{\a+2}{2} \Phi(\t_i) \f{F_{\vp j}}{F_{\vp\vp}} \ ,
\ee
and $\bar\vp$ is given in~(\ref{eq:field_canon}) with $c=1$. 
It is a straightforward task (see e.g.~\cite{Blas:2011ac, Karananas:2016kyt}) to show that in this case the ``scale-invariant'' metric becomes 
\be
\label{eq:dilatations_2}
G^{(\tilde{\chi})}_{IJ}=
\begin{pmatrix}
   \tilde F_{\vp\vp}& 0 \\
    0 & \tilde \c^2 \tilde f_{ij} 
\end{pmatrix} \ ,
\ee
with 
\be
\label{eq:dilatations_3}
 \tilde F_{\vp\vp} = F_{\vp\vp}\Phi^2 \ ,~~~\tilde f_{ij} = \l(\f{\a+2}{2}\r)^2 \Phi^2\l(F_{ij} - \f{1}{F_{\varphi\varphi}} F_{\vp i}F_{\vp j}\r) \ . 
\ee
Although the metric can certainly be cast into a block-diagonal form, the dilaton's $\tilde \c$ kinetic term~\emph{is not and cannot in general be made canonical}, unless of course $\tilde F_{\vp \vp} =1$---if this is the case, the theory is actually conformal.  To put it differently, a quick prescription to understand whether a given scale-invariant multifield theory is also conformally invariant is first to block-diagonalize the kinetic sector and then inspect if the dilaton has a canonical kinetic term.

\subsubsection{Geometric approach}

Much like in the single field considerations, the situation simplifies considerably by employing the dressing trick. Now, however, since we have more fields in the theory, we need to be a bit careful if we wish to recover the action we found before. One should allow for the $\theta_i$ fields to couple to the Ricci scalar and thus interact with the ``geometry'' in a nonminimal manner. 

In other words, the starting point should read 
\be
S = -\int d^Dx \sqrt{H}\l[ \f{\D_\vp^2}{2 (D-1)(D-2)}  M^{D-2} F_{\varphi\varphi} \mathcal R[H]  +\f 1 2 H^{\m\n} M^{D-2} F_{ij} \p_\m \t_i \p_\n\t_j + M^{D} v(\t_i)  \r] \ ,
\ee
where for obvious reasons we have denoted the nonminimal coupling function with $F_{\phi\phi}$.  Using expressions~(\ref{eq:dressed_metric}), after some trivial algebra we end up with
\be
S= -\int \diff^Dx~\Big(K +V\Big) \ ,
\ee
where the kinetic part of the action $K$ is given in~(\ref{eq:conf_K_intermediate}) and the potential $V$ in~(\ref{eq:conformalPotential}).

\subsubsection{Biscalar theory}

We shall briefly discuss the case of two scalar fields, $N=2$. Then the field-space metric \eqref{eq:field_space_metric_diagonal} is flat since it is nothing more than the metric of a two-dimensional Euclidean space expressed in polar coordinates. It can be brought to the conventional form in terms of $\phi_1,~\phi_2$:
\be
\label{eq:polar}
\c = \sqrt{\phi_1^2+\phi_2^2} \ ,~~~\t = \arctan\l(\f{\phi_2}{\phi_1}\r) \ . 
\ee
Therefore, conformality implies that the field-space curvature $\kappa$ vanishes. Conversely, a non-zero $\kappa$ implies that the field-space metric cannot be brought to the form \eqref{eq:field_space_metric_diagonal} and the absence of conformal symmetry. We conclude that 
\be \label{eq:vanishing_field_space_curvature}
\d_c S = 0 \qquad \Leftrightarrow \qquad \kappa = 0 \;.
\ee

The field-space curvature that results from a generic biscalar scale-invariant theory \eqref{eq:conf_K_start} reads 
\be \label{eq:field_space_curvature_jordan}
\kappa = \f{\left(F_{\varphi\varphi}' - (2+\alpha) F_{\varphi\t}\right)\Gamma'+2\Gamma\left((2+\alpha)F_{\varphi\t}' -  F_{\varphi\varphi}''\right)}{2\varphi^{\alpha + 2}\Gamma^2}\ ,
\ee
where we used \eq \eqref{eq:coef_dilatations_only}. Moreover, we defined $\Gamma = F_{\varphi\varphi} F_{\t\t}-F_{\varphi\t}^2$ and prime denotes differentiation with respect to $\theta$. We see explicitly that $\kappa$ vanishes when condition \eqref{eq:notindependentmixing} is fulfilled. For the  scale- but not conformally- symmetric biscalar theory, the field-space metric~\eqref{eq:dilatations_2} makes evident that curvature $\kappa$ is generically non-vanishing.

\subsection{Nonlinear realization}

Let us write down the most general $N$-field  action that we shall require to be invariant under nonlinearly realized conformal symmetry. Even when mass scales have been generated, the resulting theory retains memory of its conformally-invariant parent---this we explicitly demonstrate now.  The reason why one may be interested in that is because such situations naturally arise when gravity enters the picture and one works in the so-called Einstein frame, see the next section for a detailed discussion. 

Without loss of generality, take the action to comprise $N-1$ inert fields $\t_i$, and  one\,\footnote{When spacetime symmetries are spontaneously broken the number of NG bosons is not necessarily equal to the number of the broken generators. The textbook example of this is actually the conformal group. Although $D+1$ generator are broken---one associated with dilatations and $D$ with special conformal transformations---only one NG mode is present in the spectrum of the effective theory in the broken phase.}  Nambu-Goldstone (NG) field $\rho$ that transforms as
\be
\label{eq:shift_sym}
\d_s\rho = \p_\m \epsilon^\m  \ ,
\ee
with $\epsilon_\m$ the conformal Killing vector~(\ref{eq:conf_killing}). Therefore, our starting point reads 
\be
S = -M^2 \int \diff^D x \l [\f{\g_{\rho\rho}}{2}(\p_\m \rho)^2+ \g_{\rho i}\p_\m \rho \p^\m \theta_i +\f{\g_{ij}}{2}\p_\m \theta_i \p^\m \theta_j + M^2 U  \r] \ ,
\ee
with the  $\g$-coefficient functions and the potential $U$ depending on $\theta_i$ only. The response of the action to the transformation~(\ref{eq:shift_sym}) is straightforward to compute and  reads
\be
\d_s S = -M^2\int \diff^D x ~\p^\m \p_\n \eps^\n\l (\rho\frac{\p\g_{\rho\rho}}{\p\theta_i}  - \g_{\rho i} \r) \p_\m\theta_i \ .
\ee
Since $\g_{\rho\rho},\g_{\rho i}$ are functions of $\theta_i$ only, it can be readily verified that (up to  a total derivative) the variation of the action vanishes provided that\,\footnote{Equivalently, one may work at the level of the equations of motion for the fields
\bea
&&\g_{\rho\rho}\p^2\rho +\f{\p\g_{\rho\rho}}{\p \t_i}\p_\m \rho\p^\m\t_i  +\f{\p\g_{\rho i}}{\p \t_j}\p_\m \t_i\p^\m\t_j = 0 \ , \nonumber\\
&&
  \g_{ik}\p^2 \theta_i +\l(\f{\p\g_{ik}}{\p \theta_j} - \f{1}{2} \f{\p\g_{ij}}{\p \theta_k} \r) \p_\m \theta_i\p^\m \theta_j +\l( \f{\p\g_{\rho k}}{\p \theta_i} -\f{\p\g_{\rho i}}{\p \theta_k}  \r)\p_\m\rho\p^\m\theta_i  -\f 1 2 \f{\p\g_{\rho \rho}}{\p\theta_k}\p^2 \rho -
 \f{M^2}{2}\f{\p U}{\p \theta_k} = 0 \ .\nonumber
\eea
Invariance of the above under the shift~(\ref{eq:shift_sym}) of $\rho$ translates into
\be
\f{\p\g_{\rho\rho}}{\p\theta_i} = 0 \ ,~~~\f{\p\g_{\rho k}}{\p \theta_i} -\f{\p\g_{\rho i}}{\p \theta_k}  = 0 \ ,
\ee
which yield~(\ref{eq:grad}). 
}   
\be
\label{eq:grad}
\g_{\rho\rho}={\rm const.} \ ,~~~\g_{\rho i} = \f {\p f}{\p \theta_i} \ ,~~~{\rm with}~~~f=f(\theta_i) \ .
\ee
Consequently,  
\be
S = -M^2 \int \diff^D x \l [\f{\g_{\rho\rho}}{2}(\p_\m \rho)^2+ \f {\p f}{\p \theta_i}\p_\m \rho \p^\m \theta_i+\f{\g_{ij}}{2}\p_\m \theta_i \p^\m \theta_j + M^2 U  \r] \ ,
\ee
and a simple change of the field variable $\rho$
\be
\rho~\mapsto~\f{1}{\sqrt{\g_{\rho\rho}}}\l(\rho - \f{1}{\sqrt{\g_{\rho\rho}}}f\r) \ ,
\ee
canonicalizes its kinetic term so that  the action becomes
\be
\label{eq:action_nonlinear_final}
S = -M^2 \int \diff^D x \l [\f 1 2(\p_\m \rho)^2 +\f{\tilde\g_{ij}
}{2}\p_\m \theta_i \p^\m \theta_j + M^2 U  \r] \ ,
\ee
with $\tilde \g_{ij}$ given by
\be
\tilde \g_{ij} = \g_{ij} -\f{1}{\g_{\rho\rho}} \f {\p f}{\p \theta_i}\f {\p f}{\p \theta_j} \ .
\ee
As expected, the above is very similar to what we found in the case of the linearly realized conformal symmetry---indeed, we could start from~(\ref{eq:kin_canonical_dilaton}) and~(\ref{eq:pot_canonical_dilaton}), and expanding the dilaton on top of a nonvanishing expectation value (provided of course that the potential supports such a flat direction)
\be
\c \simeq M(1+\rho) \ ,
\ee
we end up with~(\ref{eq:action_nonlinear_final}) after appropriate identifications between the various functions appearing in the $\t$-sector. Especially in the presence of gravity, this has to be done with some care, since in general both the kinetic and potential terms  are nontrivially modified from contributions of gravitational origin. Equivalently, we can also think in terms of the field-space geometry: much like in the previous considerations, the resulting metric is block diagonal and the NG's kinetic term canonical.

We conclude by pointing out that, as expected from the results~(\ref{eq:dilatations_1})-(\ref{eq:dilatations_3}), for the case of nonlinearly realized dilatations, we would end up with a non-canonical kinetic term for $\rho$, i.e.
\be
S_{\rm dilatations}\supset -M^2 \int \diff^D x \l [\f {F(\t_i)} {2}(\p_\m \rho)^2 + \ldots \r] \ ,
\ee
with $F(\t_i)$ a function of $\t_i$. A non-canonical dilaton is a pronounced  feature of scale but not conformally invariant theories.

\section{EC scale-invariant gravity}
\label{sec:gravity}

For concreteness, we will be working in the context of the EC gravity, but our results and logic are generalizable in a straightforward manner to other formulations of gravity---what differs is the exact form of the final coefficient functions, see Eqs.~(\ref{eq:gxx})-(\ref{eq:gij}) and~(\ref{eq:gst})-(\ref{eq:gtt}) later.  
An easy computation reveals that when the metric and scalar fields are rescaled as
\be
\label{eq:scale_transformations_fields}
g_{\m\n} \to  q^{-2} g_{\m\n} \ ,~~~\phi_i(x)\to q^{\D_i}\phi_i(x) \ ,
\ee
where $q={\rm constant}$, the relevant for our discussion geometrical objects behave as
\be
\label{eq:transf_scale_quantities}
\mathring R\to q^2\mathring R \ ,~~~v_\m \to v_\m \ ,~~~a_\m \to a_\m \ ,~~~\tau_{\m\n\rho} \to q^{-2}\tau_{\m\n\rho} \ , 
\ee
with $\mathring R$ the (torsion-free) Ricci scalar constructed out of derivatives of the metric tensor $g_{\m\n}$, while $v_\m,~a_\m,~\tau_{\m\n\rho}$ are the three irreducible components of  torsion, i.e. the vector, pseudovector and reduced tensor, respectively. For a concise overview of the EC gravity basics, the interested reader is referred to~\cite{Karananas:2021zkl} and references therein.  

We will require that the purely gravitational sector of the theory is indistinguishable from the metrical GR in the absence of matter. In order to achieve this for EC gravity, it suffices to require that the admissible gravitational invariants comprise terms which are linear in curvature and at most quadratic in torsion; see discussion in \cite{Karananas:2021zkl}. Moreover, we will confine ourselves to four spacetime dimensions, since for $D\neq 4$ extra care is needed especially when decomposing torsion into its irreducible pieces by employing the totally antisymmetric symbol. 

The most general Jordan-frame action whose scalar sector comprises~(\ref{eq:kin_canonical_dilaton}) and~(\ref{eq:pot_canonical_dilaton}), while its  scale-invariant gravitational dynamics satisfy the aforementioned selection criteria reads as
\be
\begin{aligned}
\label{eq:action_conformal_torsion}
S &= \int \diff^4 x \sqrt{g} \Bigg [ \f{\s^2 G_R }{2 } \mathring R -\f 1 2 \p_\m \s \p^\m \s -\f{\s^2}{2} f_{ij}\p_\m\t_i\p^\m\t_j   -V +J^v_\m v^\m+ J^a_\m a^\m  \\
&+\f{\s^2}{2} \l( G_{vv}v_\m v^\m +2G_{va}v_\m a^\m+G_{aa}a_\m a^\m+ G_{\tau\tau}\tau_{\m\n\rho}\tau^{\m\n\rho} +\tilde G_{\tau\tau}\epsilon^{\m\n\rho\s}\tau_{\lambda\m\n}\tau^{\lambda}_{\,\rho\s}\r) \Bigg ] \ ,
\end{aligned}
\ee
where $g=-\det (g_{\m\n})$.  As usual, the contraction of spacetime indices is done with $g_{\m\n}$.  The various coefficient functions that couple torsion and curvature to the scalar fields are functions of the $\theta$'s
\be
G_R=G_R(\t_i) \ ,~~~G_{ij}=G_{ij}(\t_k) \ , 
\ee
while for dimensional reasons, the ``generalized currents'' $J^{v/a}_\m$ are 
\be
J^{v/a}_\m = \z^{v/a}_\s(\t_i)\p_\m\s^2 +\s^2 \z^{v/a}_i (\t_j)\p_\m \t_i \ .
\ee
with $\z$'s depending on $\t_i$ as explicitly shown above. We remark that we have already considered a scale-invariant coupling of two scalar fields to gravity in \cite{Karananas:2021gco}, \ie the theory \eqref{eq:action_conformal_torsion}  represents the generalization to multiple fields. 

For the subsequent analysis, we shall eliminate the nondynamical torsion from the theory; this is achieved by obtaining the equations of motion for the three torsions ($v,a,\tau$), solving them in terms of the other fields---$\s$ and $\t_i$ in the case at hand---and plugging the result back in the action~(\ref{eq:action_conformal_torsion}). We find (see also~\cite{Karananas:2021zkl,Karananas:2021gco})
\be
\begin{aligned}
S = \int \diff^4 x \sqrt{g} \Bigg [ \f{\s^2 G_R }{2 } \mathring R &-\f 1 2 \p_\m \s \p^\m \s -\f{\s^2}{2} f_{ij}\p_\m\t_i\p^\m\t_j   -V \\
&-\f{G_{aa}(J^v_\m)^2 + G_{vv}(J^a_\m)^2-2G_{va} J^v_\m J^{a\m}}{2\s^2(G_{vv}G_{aa}-G_{va}^2)}\Bigg] \ .
\end{aligned}
\ee
Expanding the above expression we obtain
\be
\begin{aligned}
S = \int \diff^4 x \sqrt{g} \Bigg [ \f{\s^2 G_R }{2 } \mathring R &-\f 1 2\l(1 + 4\f{G_{aa}(\z^v_\s)^2+G_{vv}(\z^a_\s)^2 -2G_{va}\z^v_\s \z^a_\s }{G_{vv}G_{aa}-G_{va}^2} \r) \p_\m \s \p^\m \s \\
&-2\f{G_{aa}\z^v_\s\z^v_i +G_{vv}\z^a_\s\z^a_i -G_{va}(\z^v_\s \z^a_i +\z^a_\s \z^v_i) }{G_{vv}G_{aa}-G_{va}^2}\s \p_\m \s \p^\m\t_i\\
&- \f{\s^2}{2}\l( f_{ij}+ \f{G_{aa}\z^v_i\z^v_j+G_{vv} \z^a_i\z^a_j -G_{va}(\z^v_i \z^a_j+\z^a_i \z^v_j)}{G_{vv}G_{aa}-G_{va}^2}\r) \p_\m\t_i\p^\m\t_j   -V\Bigg] \ .
\end{aligned}
\ee
We immediately notice the nontrivial torsional contributions to the kinetic terms of the fields. 

It is convenient to bring the theory in a form in which gravity is minimally coupled before we continue our analysis of scale - and conformal symmetry. To this end, we now perform the usual Weyl rescaling of the metric tensor 
\be
g_{\m\n}~\mapsto~\f {M_P^2}{\s^2 G_R} g_{\m\n} \ ,
\ee
to end up with an action possessing a canonical gravitational sector 
\be
\begin{aligned}
\label{eq:eq:action_conformal_notorsion_Einstein}
S = \int \diff^4 x \sqrt{g} \Bigg [ \f{M_P^2}{2 } \mathring R -\f {1}{2}  \tilde\g_{IJ}g^{\m\n}\p_\m \vp_I\p_\n \vp_J    -\f{M_P^4 V}{\sigma^4 G_R^2}\Bigg] \ .
\end{aligned}
\ee
We denoted with $\tilde\g_{IJ}$ the metric of the $N$-dimensional manifold spanned by $\vp_I=\l(\rho,\t_i\r)$, where we defined $\rho=\log(\s/M_P)$. Its explicit form is 
\be
 \tilde \g_{IJ}= M_P^2 \g_{IJ}  \ ,~~~ \g_{IJ} = \begin{pmatrix}
     \g_{\rho\rho} &  \g_{\rho i} \\
     \g_{\rho i} &  \g_{ij} 
\end{pmatrix} \ ,
\ee
with
\bea
\label{eq:gxx}
&&  \g_{\rho\rho} =  6 +  \f{1}{G_R}\l(1 + 4\f{G_{aa}(\z^v_\s)^2+G_{vv}(\z^a_\s)^2 -2G_{va}\z^v_\s \z^a_\s }{G_{vv}G_{aa}-G_{va}^2} \r) ,\\
&& \g_{\rho i} =  \f{2}{G_R}\l(\f{G_{aa}\z^v_\s\z^v_i +G_{vv}\z^a_\s\z^a_i -G_{va}(\z^v_\s \z^a_i +\z^a_\s \z^v_i) }{G_{vv}G_{aa}-G_{va}^2} + \f 3 2\f{\p G_R}{\p\t_i}\r) , \label{eq:gxi}\\
&& \g_{ij} = \f{1}{G_R}\l( f_{ij}+ \f{G_{aa}\z^v_i\z^v_j+G_{vv} \z^a_i\z^a_j -G_{va}(\z^v_i \z^a_j+\z^a_i \z^v_j)}{G_{vv}G_{aa}-G_{va}^2}+\f {3}{2 G_R} \f{\p G_R}{\p \t_i}\f{\p G_R}{\p \t_j} \r). 
\label{eq:gij}
\eea
Since gravity is minimally coupled in \eq \eqref{eq:eq:action_conformal_notorsion_Einstein}, we can now return to analysing conformality. Namely, one could require that this symmetry is exclusively broken by the dynamical graviton, \ie that the action exhibits exact non-linearly realised conformal invariance in the limit $\mathring R\rightarrow 0$. As derived in \eq \eqref{eq:grad}, this amount to demanding\,\footnote {In general, there is no relationship -- not even an approximate one -- between the requirements \eqref{eq:conformal_gravity_torsion_einstein1} and \eqref{eq:conformal_gravity_torsion_einstein2} derived in the Einstein frame and the conditions 
\be 
\label{eq:conformal_gravity_torsion}
\f{G_{aa}\z^v_\s\z^v_i +G_{vv}\z^a_\s\z^a_i -G_{va}(\z^v_\s \z^a_i +\z^a_\s \z^v_i) }{G_{vv}G_{aa}-G_{va}^2} = \f{\p}{\p \t_i}\l(\f{G_{aa}(\z^v_\s)^2+G_{vv}(\z^a_\s)^2 -2G_{va}\z^v_\s \z^a_\s }{G_{vv}G_{aa}-G_{va}^2} \r)\ , 
\ee
corresponding to  Jordan frame conformality, as follow from~(\ref{eq:notindependentmixing}).  For example, one can specialize to the case of two fields and consider $G_R=1$. In this case, the only non-trivial condition that follows from imposing exact conformal invariance in the Einstein frame consists in \eq \eqref{eq:conformal_gravity_torsion_einstein1}. From the Jordan frame perspective, this implies that the \rhs of \eq \eqref{eq:conformal_gravity_torsion} vanishes. However, the \lhs (in particular the couplings $\z^v_\t$ and $\z^a_\t$) remains completely unconstrained. Therefore, such a theory does not fulfil -- even approximately -- the Jordan frame condition \eqref{eq:conformal_gravity_torsion} (see also \eq \eqref{eq:field_space_curvature_jordan}).}
\begin{align}
\frac{\p}{\p_{\theta_i}}\l(  \f{1}{G_R}\l(1 + 4\f{G_{aa}(\z^v_\s)^2+G_{vv}(\z^a_\s)^2 -2G_{va}\z^v_\s \z^a_\s }{G_{vv}G_{aa}-G_{va}^2} \r)\r) &=0\;, \label{eq:conformal_gravity_torsion_einstein1}\\
 \f{2}{G_R}\l(\f{G_{aa}\z^v_\s\z^v_i +G_{vv}\z^a_\s\z^a_i -G_{va}(\z^v_\s \z^a_i +\z^a_\s \z^v_i) }{G_{vv}G_{aa}-G_{va}^2} + \f 3 2\f{\p G_R}{\p\t_i}\r) &= \frac{\p}{\p_{\theta_i}} f \ ,~~~{\rm with}~~~f=f(\theta_i) \;. \label{eq:conformal_gravity_torsion_einstein2}
\end{align}
However, we shall not impose conditions \eqref{eq:conformal_gravity_torsion_einstein1} and \eqref{eq:conformal_gravity_torsion_einstein2} in the following. Since gravity breaks conformality in any case, imposing exact conformality is too strong a requirement.

Before we further discuss this point, we will turn to the phenomenologically interesting situation comprising $N=2$ scalar fields. In this case, we get a two-dimensional target manifold with metric 
\be
\g_{IJ} =   \begin{pmatrix}
     \g_{\rho\rho} &  \g_{\rho \t} \\
    \g_{\rho \t} &  \g_{\t\t} 
\end{pmatrix} \ ,
\ee
where $\g_{\s\s}$ is given by~(\ref{eq:gxx}), while (prime stands for differentiation w.r.t. $\t$) 
\bea
\label{eq:gst}
&& \g_{\rho\t} = \f{2}{G_R}\l(\f{G_{aa}\z^v_\s\z^v_\t +G_{vv}\z^a_\s\z^a_\t-G_{va}(\z^v_\s \z^a_\t+\z^a_\s \z^v_\t) }{G_{vv}G_{aa}-G_{va}^2} + \f 3  2 G_R'\r) , \label{eq:gxt}\\
&& \g_{\t\t} = \f{1}{G_R}\l( f_{\t\t}+ \f{G_{aa}(\z^v_\t)^2+G_{vv} (\z^a_\t)^2 -2G_{va}\z^v_\t \z^a_\t}{G_{vv}G_{aa}-G_{va}^2}+\f {3}{2 G_R} G_R'^2 \r) \ , \label{eq:gtt}
\eea
and the various coefficient functions now only depend on $\t$. It is clear that there is too much arbitrariness in the functions, translating into the theory (in the absence of gravity, i.e. for $\mathring R\to 0$) not exhibiting spontaneously broken conformal symmetry.

 As a consistency check, we can momentarily restrict ourselves to the scale-invariant Higgs-dilaton theory considered in \cite{Karananas:2021gco}:
\bea  \label{actionGeneral}
	S &&= \int \diff^4 x \sqrt{g}\Bigg[\frac{1}{2}\left(\x_1 \phi_1^2 + \x_2 \phi_2^2\right)\mathring{R} - \frac {(\partial_\mu \phi_1)^2}{2} -  \frac {(\partial_\mu \phi_2)^2}{2} - \frac{\lambda}{4} \phi_1^4\nonumber\\
	&& +  \left(\zeta^{v}_1 \partial_\mu \phi_1^2 + \zeta^{v}_2 \partial_\mu \phi_2^2\right) v^\mu+   \left(\zeta^{a}_1 \partial_\mu \phi_1^2 + \zeta^{a}_2 \partial_\mu \phi_2^2\right) a^\mu \\
	&&+  \frac{c_{vv}}{2} \left(\xi_{vv}\phi_1^2+\xi_2 \phi_2^2 \right) v_\mu v^\mu + c_{va}\left(\xi_{va}\phi_1^2+\xi_2 \phi_2^2 \right) v_\mu a^\mu +  \frac{c_{aa}}{2} \left(\xi_{aa}\phi_1^2+\xi_2 \phi_2^2 \right) a_\mu a^\mu \Bigg] \ . \nonumber
\eea
Here $\phi_1$ and $\phi_2$ correspond to the Higgs field in unitary gauge and dilaton, respectively. Moreover, $\x_1,\x_2,\x_{ij}, c_{ij},\z^{v/a}_1,\z^{v/a}_2,\lambda$ are real constants (with $i,j=a,v$) and we can recover the polar variables $\s, \t$~(see eq.~(\ref{eq:polar})) with the field redefinition
\be \label{eq:coordinate_transformation}
\phi_1 = \s \cos\t \ ,~~~\phi_2 = \s \sin\t \ .
\ee
In \eq \eqref{eq:action_conformal_torsion}, the parametrization of \eq \eqref{actionGeneral} corresponds to setting 
\be
\begin{aligned}
\label{eq:TDFs_HD}
&G_R  = \x_1 \cos^2\t +\x_2 \sin^2\t \ ,~~~f_{\t\t}=1 \ ,~~~G_{ij} = c_{ij}\l(\x_{ij} \cos^2\t +\x_2 \sin^2\t \r) \ ,\\
&\z^{v/a}_\s = \z_1^{v/a}\cos^2\t +\z_2^{v/a}\sin^2\t \ ,~~~\z^{v/a}_\t = \l(\z_2^{v/a}-\z_1^{v/a}\r)\sin 2\t \ ,~~~V=\f\lambda 4 \s^4 \cos^4\t\ ,
\end{aligned}
\ee
where no summation over the repeated Latin indices is assumed in~(\ref{eq:TDFs_HD}). 
Plugging the transformation \eqref{eq:coordinate_transformation} in the action \eqref{eq:eq:action_conformal_notorsion_Einstein} leads to 
\be
\begin{aligned}
S = \int \diff^4 x \sqrt{g} \Bigg [ \f{M_P^2}{2 } \mathring R &-\f {1}{2\s^2 G_R}  \g^{\phi_1,\phi_2}_{IJ}g^{\m\n}\p_\m \phi_I\p_\n \phi_J    -\f{M_P^4V}{ \sigma^4 G_R^2}\Bigg] \ ,
\end{aligned}
\ee
where
\bea
&&   \g^{\phi_1,\phi_2}_{11} = \f{G_R M_P^2}{\s^2} \left(\phi_1^2    \g_{\rho\rho} - 2 \phi_1 \phi_2   \g_{\rho\t} + \phi_2^2    \g_{\t\t}\right) \  ,\\
&&  \g^{\phi_1,\phi_2}_{12} = \f{G_R M_P^2}{\s^2} \left(\phi_1 \phi_2    \g_{\rho\rho} + \left(\phi_1^2-\phi_2^2\right)   \g_{\rho\t} - \phi_1 \phi_2   \g_{\t\t}\right)  \ ,\\
&&  \g^{\phi_1,\phi_2}_{22} = \f{G_R M_P^2}{\s^2} \left(\phi_2^2    \g_{\rho\rho} + 2\phi_1 \phi_2   \g_{\rho\t} + \phi_1^2    \g_{\t\t}\right) \ .
\eea
Inserting \eqs \eqref{eq:gxx}, \eqref{eq:gxt} and \eqref{eq:gtt} into the above, we obtain \eq (26) of \cite{Karananas:2021gco}.\footnote{In order to reproduce the notation of \cite{Karananas:2021gco}, one has to replace $\s^2 G_R\rightarrow \Omega^2$ and $G_{ij}\rightarrow G_{ij} \xi_2 \phi_2^2/\s^2 $.}

\subsection{Conformal symmetry up to the Planck scale}

So far, we have derived a general class of scale-invariant theories described by \eqs \eqref{eq:gxx}, \eqref{eq:gxi} and \eqref{eq:gij}. In the case of two scalar fields, the latter two equations are replaced by \eqs \eqref{eq:gxt} and \eqref{eq:gtt}.
Now we turn to conformal symmetry. As discussed before, it cannot be preserved in the presence of gravity. As long as coupling to gravity is minimal, however, conformal invariance is only broken at the Planck scale $M_P$. At lower energies, effects of dynamical gravity are suppressed and one can achieve approximate conformality in analogy to the state of affairs in flat spacetime. 

As shown above, this situation changes once scalar matter couples to gravity non-minimally. Then effects violating conformal invariance can appear far below $M_P$. Our goal is to keep the violation of conformality ``minimal,'' even if non-minimal couplings are included. This amounts to imposing that the scalar sector of the theory exhibits approximate conformal invariance up to the Planck scale $M_P$. In other words, effects that break conformal symmetry should be suppressed by the Planck scale. We can express this criterion in terms of field-space curvature $\kappa$. Since $\kappa=0$ corresponds to exact conformal invariance (see \eq \eqref{eq:vanishing_field_space_curvature}), we now require\,\footnote
{To make this point clear, consider the following toy model of two real scalar fields $\phi_1$ and $\phi_2$ with canonical mass dimension and action
\be
S = \int \diff^4 x \sqrt{g} \Bigg [-\f {1}{2}  \p_\m \phi_1\p^\m \phi_1  -\f {1}{2}\l( 1+ \f{\tilde{\kappa} \phi_1^2}{2M_P^2}\r)  \p_\m \phi_2\p^\m \phi_2\Bigg] \ .
\ee
The higher-dimensional operator suppressed by the scale $M_P/\sqrt{\tilde{\kappa}}$ explicitly breaks conformal invariance---in its absence the above simply comprises two free canonical scalar fields. The field-space curvature in the limit $\phi_1\rightarrow 0$ is $|\kappa| = \tilde{\kappa}/M_P^2$; demanding $|\kappa|\lesssim 1/M_P^2$ indeed implies $\tilde{\kappa}\lesssim 1$, \ie that the scale of conformality breaking is in the vicinity of $M_P$.} 
\be
\label{eq:curv_planck}
|\kappa| = \left\vert\f{\g_{\rho\rho}'\Gamma'-2\Gamma  \g_{\rho\rho}''}{2M_P^2 \Gamma^2}\right\vert \lesssim \frac{1}{M_P^2} \ ,
\ee
with $\Gamma =  \g_{\rho\rho}  \g_{\t\t}-\g_{\rho\t}^2$.
Demanding that this condition holds for all values of the scalar fields significantly constrains the parameters of the theory and in this way substantially reduces the ambiguity that results from the different formulations of GR.

In \eqs \eqref{eq:conformal_gravity_torsion_einstein1} and \eqref{eq:conformal_gravity_torsion_einstein2}, we had already discussed a requirement of exact conformal invariance in the limit $\mathring R\to 0$ of non-dynamical gravity. Imposing this is stronger than the bound \eqref{eq:curv_planck} on field space curvature: When \eqs \eqref{eq:conformal_gravity_torsion_einstein1} is fulfilled, then field-space curvature vanishes identically, $\kappa = 0$.

\section{``Higgs-type Inflation''? A conjecture.}
\label{sec:higgs_inflation}

 As discussed in \cite{Shaposhnikov:2020gts,Langvik:2020nrs}, numerous a priori unknown coupling constants exist in EC gravity and other formulations of GR (see \eg \cite{Heisenberg:2018vsk, BeltranJimenez:2019esp, Rigouzzo:2022yan} for an overview). This leads to a built-in ambiguity when it comes to inflating with the  Higgs field, since the observable predictions are not unique but depend on the gravitational incarnation. The reason why this is so can be understood by noting that choosing to work in the context of a particular formulation of gravity translates into choosing a specific set of higher-dimensional operators when the theory is written in its equivalent purely metrical form. In turn, these operators feed into the kinetic term for the field shaping its behaviour in specific manners. 

Clearly, imposing conformal symmetry up to the Planck scale improves this situation since this requirement constrains the a priori unknown coupling constants and reduces the arbitrariness that exists in EC gravity. We conjecture that the situation may be even better,   at least for the biscalar theories, and that a much stronger statement could hold:

\textbf{Conjecture:} In the Einstein-Cartan formulation of General Relativity, we consider the Higgs-dilaton model~\eqref{actionGeneral}, which is constructed according to the criteria of \cite{Karananas:2021zkl} so that the  coefficient functions comprise operators polynomial in the two fields and of mass dimension at most four. If the high-energy value of the Higgs self-coupling fulfills $\lambda \gtrsim 10^{-12}$ and 
\begin{enumerate}
    \item conformality is  preserved up to the Planck scale, \ie $\kappa \lesssim 1/M_P^2$ for all field values,
    \item slow-roll inflation is possible,
    \item the observed amplitude of CMB perturbations is reproduced,
\end{enumerate}
then the inflationary predictions fulfill (to leading order in $1/N$)
\be \label{conjecture}
n_s = 1- \f 2 N \ ,~~~r \gtrsim \f{12}{N^2}  \;,
\ee
where $n_s$ is the spectral index, $r$ the tensor-to-scalar ratio and $N$ corresponds to the number of e-foldings between the generation of CMB and the end of inflation.

If we had an equality in the second equation of our conjecture \eqref{conjecture}, \ie $r=12/N^2$, then this would coincide with the predictions of single-field Higgs inflation in the metric formulation of GR \cite{Bezrukov:2007ep}. Thus, our conjecture implies that the requirement of conformality brings the generic Higgs-dilaton model \eqref{actionGeneral} with its numerous unknown parameters close to the scenario of metric Higgs inflation, in which only one coupling constant is added to the ones already present in the Standard Model:\footnote
{In a different setting, a connection of conformal invariance and predictions of single-field metric Higgs inflation was also pointed out in \cite{Clery:2023ptm}.} the spectral indices are identical and the tensor-to-scalar ratio is bounded from below by the value derived from metric Higgs inflation.
Finally, we remark that the condition on $\lambda$ is very mild since typical values are around $\lambda \sim 10^{-3}$ (see \cite{Shaposhnikov:2020fdv}).\footnote
{If the high-energy value of the self-coupling were as small as $\lambda \sim 10^{-13}$, then it would become possible to implement Higgs inflation without any non-minimal coupling. However, the predictions derived from an Einstein-frame potential $\lambda h^4$ do not match CMB obervations \cite{Akrami:2018odb, BICEP:2021xfz}.}

We have three motivations for our conjecture. The first one comes from the Higgs-dilaton model in the metric formulation of GR \cite{Garcia-Bellido:2011kqb}. In this case, certain parts of the parameter space lead to a field-space curvature that approaches Planckian values, $\kappa \lesssim 1/M_P^2$, in the limit of a large Higgs field. For such choices of coupling constants, the predictions coincide with the ones of single-field metric Higgs inflation.\footnote
{Even though the field-space curvature of the metrical Higgs-dilaton model, which is defined in \eq \eqref{metricHiggsDilaton} below, is roughly equal to the Planck area during inflation, at low energies  conformality is violated well below $M_P$, thus this specific example fails to comply with the first requirement of our conjecture. Nevertheless, its observables saturate~(\ref{conjecture}), as long as the dilaton nonminimal coupling satisfies $\x_2 < 4\times 10^{-3}$.}
Our second inspiration originates from the observation that in a certain subclass of said biscalar theories, inflationary observables are directly related to geometry. For intervals of approximately constant $\kappa$, the spectral index assumes the universal value $1-2/N$ and the tensor-to-scalar ratio can be computed as follows \cite{Karananas:2016kyt}
\be \label{rFromCurvature}
r = \f{4}{M_P^2\l|\kappa\r| N^2} \;.
\ee
Therefore, the small $|\kappa|$ of our conjecture leads to an upper bound on $r$ as in \eq \eqref{conjecture}. Finally, our third motivation is purely empirical. Attempting to construct models that obey the requirements of the  conjecture, the only examples we found fulfilled \eq \eqref{conjecture}.

\subsection{Reminder of single-field metric Higgs inflation}

For comparison, we shall briefly review metric Higgs inflation in the absence of a dilaton \cite{Bezrukov:2007ep}. The Jordan frame action is given by  
\be \label{actionHiggsMetric}
S = \int\diff^4 x \sqrt{g}\left[ \f{M_P^2}{2}\l(1 + \frac{\xi_1 h^2}{M_P^2}\r) \mathring R - \f{1}{2} (\p_\m h)^2 -\f{\lambda}{4}h^4 \right] \ ,
\ee
where $h$ is the Higgs field in unitary gauge.
For $\xi_1\gg1$ and $h^2\gg M_P^2/\x_1$, the theory in the Einstein frame reads
\be \label{actionHiggsMetricEinstein}
S \approx \int\diff^4 x \sqrt{g}\left[ \f{M_P^2}{2} \mathring R - \f{3 M_P^2}{h^2} (\p_\m h)^2 -\f{\lambda M_P^4}{4 \x_1^2} \l(1-\f{M_P^2}{\x_1 h^2}\r)^2 \right] \ .
\ee

A standard analysis of inflationary dynamics shows that to leading order in $1/N$ the inflationary indices are \cite{Bezrukov:2007ep} (\cf \eq \eqref{conjecture})
\be \label{predictionsMetric}
n_s = 1- \f 2 N \ ,~~~r = \f {12} {N^2} \ .
\ee

\subsection{Metric Higgs-dilaton}
\label{sec:metric_HD}

We shall come back to the case of two fields. The Higgs-dilaton model in the metric formulation~\cite{Shaposhnikov:2008xb,Garcia-Bellido:2011kqb} is obtained as a special case when torsion vanishes, \ie
\be
c_{ij} = \z_1^{v/a}=\z_2^{v/a}=0 \ .    
\ee
Then the action \eqref{actionGeneral} becomes
\be \label{metricHiggsDilaton}
S = \int \diff^4 x \sqrt{g}\Bigg[\frac{1}{2}\left(\x_1 \phi_1^2 + \x_2 \phi_2^2\right)\mathring{R} - \frac {(\partial_\mu \phi_1)^2}{2} -  \frac {(\partial_\mu \phi_2)^2}{2} - \frac{\lambda}{4} \phi_1^4 \Bigg] \ ,
\ee
and equivalently \eq \eqref{eq:action_conformal_torsion} yields
\be
S = \int \diff^4 x \sqrt{g}\Bigg[\frac{\s^2}{2}\left(\x_1 \cos^2\t +\x_2 \sin^2\t\right)\mathring{R} - \frac {(\partial_\mu \s)^2}{2} -  \f{\s^2}{2}(\partial_\mu \t)^2 - \frac{\lambda}{4} \s^4\cos^4\t \Bigg] \ .
\ee
Then the Einstein-frame action is given by \eq \eqref{eq:eq:action_conformal_notorsion_Einstein}, after defining $\rho=\log(\s/M_P)$ as before. As is evident from \eqs \eqref{eq:gxx} to \eqref{eq:gij}, the components of the field-space metric read
\bea
&&\g_{\rho\rho} = \f{1}{\x_1 \cos^2\t +\x_2 \sin^2\t}+6 \ , \label{eq:gxxH}\\
&&\g_{\rho\t}=  -3(\x_1-\x_2)\f{\sin2\t}{\x_1 \cos^2\t +\x_2 \sin^2\t} \ ,\\
&&\g_{\t\t} = \f{1}{\x_1 \cos^2\t +\x_2 \sin^2\t}\l(1+\f 3 2 (\x_1-\x_2)^2 \f{\sin^2 2\t}{\x_1 \cos^2\t +\x_2 \sin^2\t}\r) \ , \label{eq:gijH}
\eea
where we used \eq \eqref{eq:TDFs_HD}. 

 For the subsequent analysis, it is useful to eliminate the kinetic mixing between the fields. This can be achieved by shifting $\rho$ as~\cite{Blas:2011ac,Karananas:2016kyt,Karananas:2021gco} 
\be
\label{eq:tilderho}
\rho \to \tilde \rho = \rho + \int \diff\t \f{\g_{\rho\t}}{\g_{\rho\rho}}  \ ,
\ee
which in turn translates into the action becoming 
\be 
S = \int\diff^4 x \sqrt{g}\left[ \f{M_P^2}{2} \mathring R - \f{M_P^2}{2} \l(\g_{\rho\rho} (\p_\m \tilde\rho)^2+\mathcal K (\p_\m \t)^2\r) -\f{\lambda M_P^4}{4\x_1^2} \l(\f{\x_1}{\x_1 + \x_2\tan^2 \t}\r)^2  \right] \ ,
\ee
where
\be \label{Kprototype}
\mathcal K = \f{\g_{\t\t}\g_{\rho\rho}-\g_{\rho\t}^2}{\g_{\rho\rho}} \ .
\ee

 In order to get a grasp on the field-space geometry, it is convenient to perform another change of variables. Introducing 
 \be
 Z=  \f{\x_2\tan^2\t}{\x_1+\x_2 \tan^2\t}\ , 
\ee
we get
\be
\label{eq:actionHD_Z}
S=\int\diff^4 x \sqrt{g}\left[ \f{M_P^2}{2} \mathring R - \f{M_P^2}{2}\l(\g_{\rho\rho} (\p_\m \tilde\rho)^2+\mathcal{K}_{Z} (\p_\m Z)^2\r)  -\f{\lambda M_P^4}{4 \x_1^2} \l(1-Z\r)^2 \right] \ ,
\ee
where in terms of $Z$ 
\be
\g_{\rho\rho} = \f{1}{\x_1}+ 6+\l(\f{\x_1-\x_2}{\x_1\x_2}\r)Z \ ,~~~\mathcal{K}_{Z}= \f{1+6\x_1-6(\x_1-\x_2)Z}{4Z(\x_2+6\x_1\x_2+(\x_1-\x_2)Z)(1-Z)} \ .
\ee
A straightforward computation reveals that the associated Einstein-frame field-space curvature $\kappa$ is given by
\be
M_P^2\kappa = -\f{1}{3}\l(1-\f{(1+6\x_1)(1+6\x_2)}{(1+6\x_1-6(\x_1-\x_2)Z)^2}\r) \ .
\ee

Inflation takes place for $Z \ll 1$, meaning that 
\be
M_P^2 \kappa_{\rm infl} \simeq -\f{1}{3} \ , 
\ee
while around the electroweak vacuum $Z \simeq 1$ and
\be
M_P^2 \kappa_{\rm EW} \simeq \frac{2 \x_1}{(1+6\x_2)} \ ,
\ee
where in deriving the asymptotic values we assumed $\x_1\gg 1$ and $\x_1\gg \x_2$. We observe that   during inflation the curvature is of the order of the Planck area, however,  at low energies it becomes significantly larger---this in turn translates into  $\Lambda_{\rm conf}\simeq M_P/\sqrt{\x_1}$ which is well below $M_P$ and thus fails to conform with~(\ref{eq:curv_planck}). For completeness, we note that the low-energy cutoff of the  Higgs-dilaton model is~\cite{Bezrukov:2012hx} $\Lambda_{\rm cutoff}\simeq M_P/\x_1 < \Lambda_{\rm conf}.$

To remedy the situation with conformality-breaking, we can make a different choice of non-minimal couplings, namely $\x_1=\x_2$. Then it follows from \eqs \eqref{eq:gxxH} to \eqref{eq:gijH} that the Einstein-frame action becomes
\bea
\label{eq:HD_metric_hilltop}
S = \int \diff^4 x \sqrt{g} \Bigg[\frac{M_P^2}{2}\mathring{R} - \f {M_P^2}{2\s^2}\l(\f{1}{\x_1 }+6\r)\l(\partial_\mu \rho\r)^2 -\f{M_P^2}{2\x_1 }\l(\partial_\mu \t\r)^2-\f{\lambda M_P^4}{4\x_1^2}\cos^4\t \Bigg].
\eea
Since $\g_{\rho\rho}={\rm constant}$ and $\g_{\rho\t}=0$, the target manifold is flat at all field values:
\be
\kappa = 0 \;.
\ee
    Thus, the model \eqref{eq:HD_metric_hilltop} fulfills requirement \eqref{eq:curv_planck}. It remains to be checked, however, if inflation can be realized. To this end, we introduce a canonically normalized angular field 
\be
\tilde \t = \f{M_P}{\sqrt{\x_1}}\l(\t +\f \pi 2 \r) \ ,
\ee
where we additionally shifted $\t$ such that the potential has its minimum at vanishing field values. Then the part of the action \eqref{eq:HD_metric_hilltop} that is relevant for inflation reads
\bea
\label{eq:HD_metric_hilltopCan}
S = \int \diff^4 x \sqrt{g} \Bigg[\frac{M_P^2}{2}\mathring{R} -\f{1}{2}\l(\partial_\mu \tilde\t\r)^2-\f{\lambda M_P^4}{4\x_1^2}\sin^4\l(\f{\sqrt{\x_1}\tilde \t}{M_P} \r) \Bigg].
\eea
In the limit $\x_1 \rightarrow 0$, the potential becomes $\lambda \tilde \t^4/4$. As is well-known (see \eg \cite{Martin:2013tda}), this model cannot match the observed amplitude of perturbations in the CMB unless $\lambda \sim 10^{-13}$.

 We shall now show that $\x_1=\x_2$ cannot be much bigger than $1$. To this end, we first compute the inflationary indices
\begin{align}
	\epsilon & = 8 \x_1 \frac{1}{\tan^2\left(\f{\sqrt{\x_1}\tilde \t}{M_P}\right)} \;,\\
	\eta & = 4  \x_1 \left(\frac{3}{\tan^2\left(\f{\sqrt{\x_1}\tilde \t}{M_P}\right)} - 1 \right) \ .
\end{align}
We see that the requirements $\epsilon \ll 1$ and $|\eta|\ll 1$ can only be fulfilled simultaneously if
\begin{equation}
	\x_1 \lesssim 1 \;.
\end{equation}
However, we shall not employ this -- or any other -- approximation in the following. Next, the equations $\epsilon\approx 1$ and $|\eta| \approx 1$ lead to similar conditions, and so the end of inflation occurs around
\begin{equation}
	\theta_{\text{end}} \simeq \f{M_P}{\sqrt{\x_1}} \arctan \sqrt{\x_1}\ .
\end{equation} 
We did not show numerical factors of order $1$ and as always they depend on the precise definition of the end of inflation.

Evaluating the number $N$ of inflationary e-foldings, we obtain
\be
	N = \frac{1}{4 \sqrt{\x_1} M_P} \limitint_{\tilde \t_{\text{end}}}^{\tilde \t_\star} \diff \tilde \t \,   \tan\left(\f{\sqrt{\x_1}\tilde \t}{M_P}\right) = \frac{1}{4 \xi_1} \log \left(\frac{\cos\left(\f{\sqrt{\x_1}\tilde \t_{\text{end}}}{M_P}\right)}{\cos\left(\f{\sqrt{\x_1}\tilde \t_\star}{M_P}\right)}\right) \ ,
 \ee
 meaning that the horizon exit takes place for
 \be
 \tilde \t_\star  = \f{M_P}{\sqrt{\x_1}} \arccos \left(\exp\left(- 4 N \xi_1\right) \cos\left(\f{\sqrt{\x_1}\tilde \t_{\text{end}}}{M_P}\right) \right) \ .
\ee
Thus, the slow-roll parameters evaluated on $\tilde \t_\star$ read
\begin{align}
	\epsilon & =  8 \xi_1 \frac{1}{\exp\left( 8 N \x_1\right) \cos^{-2}\left(\f{\sqrt{\x_1}\tilde \t_{\text{end}}}{M_P}\right) - 1} \;,\\
	\eta & = 4  \xi_1 \left(\frac{3}{\exp\left( 8 N \x_1\right) \cos^{-2}\left(\f{\sqrt{\x_1}\tilde \t_{\text{end}}}{M_P}\right) - 1} -1\right) \;.
\end{align}
Finally, we can evaluate the amplitude of perturbations 
\begin{equation} \label{amplitudeHilltop}
	\frac{U}{\epsilon} = \frac{\lambda M_P^4}{32 \x_1^3} \frac{\left(1- \exp\left(- 8 N \x_1\right)\cos^2\left(\f{\sqrt{\x_1}\tilde \t_{\text{end}}}{M_P}\right) \right)^3}{\exp\left(- 8 N \x_1\right) \cos^2\left(\f{\sqrt{\x_1}\tilde \t_{\text{end}}}{M_P}\right)} \;.
\end{equation}
Since $\x_1 \lesssim 1$, we can expect that generically the amplitude of perturbations is too large, \ie it is hard to fulfill the condition $U/\epsilon = 5 \cdot 10^{-7} M_P^4$. In an attempt to make $U/\epsilon$ small, we can choose the parameters such that $\exp\left(- 8 N \x_1\right) \approx 1$, or equivalently
\begin{equation}
	8N \x_1\lesssim 1 \;.
\end{equation}
Additionally using that $\cos \left(\f{\sqrt{\x_1}\tilde \t_{\text{end}}}{M_P}\right)\approx 1$, our result reduces to
\begin{align} \label{amplitudeHilltopEv}
	\frac{U}{\epsilon} & =16 \lambda N^3 M_P^4 \;.
\end{align}
This is identical to what we would have obtained in pure quartic inflation, \ie for $\x_1\rightarrow 0$. Thus, we cannot match the amplitude of observed perturbations in the CMB unless $\lambda\sim 10^{-13}$.

In summary, we have illustrated that in the metric formulation of GR it is hard to obtain a scenario of inflation that agrees with CMB and at the same time preserves conformality up until  the Planck scale. The parameter choice $\x_1\gg \x_2$ violates the latter condition whereas the model with $\x_1=\x_2$ is incompatible with the former.

\subsection{Higgs-dilaton beyond the metric formulation}

Finally, we shall present a prototype model that fulfills all requirements of our above conjecture; its action reads 
\begin{align}
	S &= \int \diff^4 x \sqrt{g}\Bigg[\frac{1}{2}\left(\xi_1 \phi_1^2 + \f{1}{\x_1^2} \phi_2^2\right)\mathring{R} - \frac {(\partial_\mu \phi_1)^2}{2} -  \frac {(\partial_\mu \phi_2)^2}{2} +   \partial_\mu \phi_2^2 v^\mu+  (\xi_1 \partial_\mu \phi_1^2 + \partial_\mu \phi_2^2) a^\mu \nonumber\\
	&+  \f{1}{2\x_1}\left(\frac{8 \xi_1^3}{\xi_1-2} \phi_1^2+ \f{\phi_2^2}{\x_1^2}  \right) v_\mu v^\mu  + \f{1}{\x_1} \left(\f{1}{\x_1}\phi_1^2+\frac{\phi_2^2}{  \xi_1^2}    \right)v_\mu a^\mu + \f{1}{2\x_1} \left(\f{1}{\x_1}\phi_1^2+\frac{\phi_2^2}{  \xi_1^2}    \right)	a_\mu a^\mu - \frac{\lambda}{4} \phi_1^4  \Bigg]\ . \label{prototype}
\end{align}
A direct comparison with~(\ref{actionGeneral}), reveals that the above corresponds to choosing
\begin{equation*}
\z_1^v = 0 \ ,~~\z_2^v=\z^a_2 =1 \ ,~~\sqrt{\x_2}=\x_{va} =\x_{aa}=\f{1}{\z_1^a}= c_{vv} =c_{va}=c_{aa}= \f{1}{\x_1}\ ,~~\x_{vv} = \f{8\x_1^3}{\x_1-2} \  . 
\end{equation*}
In terms of the angular variable $\theta$  the various theory-defining coefficient functions become
\be
\begin{aligned}
&G_R  = \x_1 \cos^2\t +\f{1}{\x_1^2}\sin^2\t \ ,~~~f_{\t\t}=1 \ ,\quad G_{vv} = \f {1}{\x_1} \left(\f{8\x_1^3}{\x_1-2} \cos^2\theta+\f{1}{\x_1^2} \sin^2 \theta  \right) \ , \\
& G_{va} = G_{aa} =  \f {1}{\x_1} \left(\f {1}{\x_1}\cos^2\theta+\f{1}{\x_1^2} \sin^2 \theta \right) \ ,\quad\z^v_\s = \sin^2\t \ , \quad \z^v_\t = \sin2\t \ , \\
&\z^a_\s = \x_1 \cos^2\t +\sin^2\t \ , \quad \z^a_\t = -(\x_1-1)\sin2\t \ ,
\end{aligned}
\ee
In turn, the Einstein-frame action is given by~(\ref{eq:eq:action_conformal_notorsion_Einstein}), where $\g_{\rho\rho}$, $\g_{\rho\theta}$ and $\g_{\theta\theta}$ are determined by \eqs \eqref{eq:gxx} to \eqref{eq:gij}.

Following closely what we did in the previous section, we first diagonalize the kinetic sector by introducing $\tilde\rho$ as in Eq.~(\ref{eq:tilderho}); we obtain
\be \label{diagonalPrototype}
S = \int\diff^4 x \sqrt{g}\left[ \f{M_P^2}{2} \mathring R - \f{M_P^2}{2} \l(\g_{\rho\rho} (\p_\m \tilde\rho)^2+\mathcal K (\p_\m \t)^2\r) -\f{\lambda M_P^4}{4\x_1^2} \l(\f{\x_1^3}{\x_1^3 + \tan^2 \t}\r)^2  \right] \ ,
\ee
where $\mathcal K$ can be explicitly found by using the expression~(\ref{Kprototype}). 
Defining $\tilde{h}^2 = M_P^2 (\tan^2 \theta + \xi_1^3)/(\xi_1 \tan^2 \theta)$, we can bring \eq \eqref{diagonalPrototype} to the form
\be \label{actionhTilde}
S = \int\diff^4 x \sqrt{g}\left[ \f{M_P^2}{2} \mathring R - \f{M_P^2}{2} \l(\g_{\rho\rho} (\p_\m \tilde\rho)^2+\mathcal{K}_{\tilde{h}} (\p_\m \tilde{h})^2\r)  -\f{\lambda M_P^4}{4 \x_1^2} \l(1-\f{M_P^2}{\x_1 \tilde{h}^2}\r)^2 \right] \ ,
\ee
where $\mathcal{K}_{\tilde{h}} = (\f{\diff \theta}{\diff \tilde{h}})^2 \mathcal{K}$.

\begin{figure}[!t]
    \centering
    \includegraphics[scale=.3]{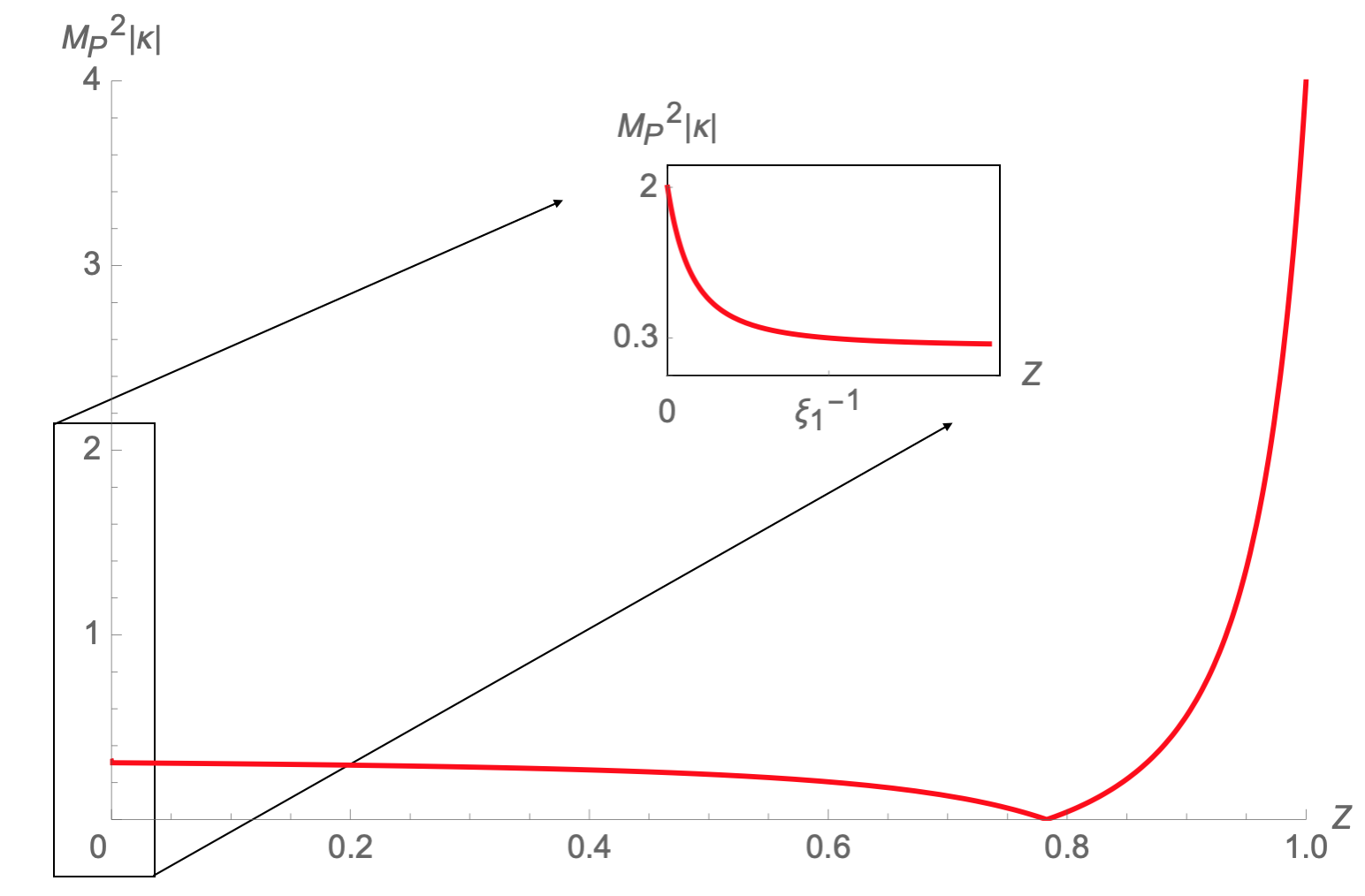}
    \caption{ The absolute value of the field-manifold curvature $M_P^2|\kappa|$ for $\x_1=10^4$ as a function of the inflaton field $Z$. We notice that its value for $Z\to 0$ is $M_P^2|\kappa| \sim2$. For non-vanishing values of the field it rapidly decays and becomes approximately constant and equal to $M_P^2|\kappa| \sim 4/13$. Inflation takes place at field values $Z\sim \x_1^{-1}$, meaning that the tensor-to-scalar ratio is expected to depend on the ``equilibrated'' value of the field-space curvature.} 
    \label{fig:curv_Z}
\end{figure}

First, we analyze inflation. To this end, we leave out $\tilde{\rho}$ since it decouples from inflationary dynamics, the latter being for all practical purposes effectively single-field~\cite{Shaposhnikov:2008xb,Garcia-Bellido:2011kqb,Karananas:2016kyt,Casas:2018fum}.
For large $\xi_1$, the coefficient of the kinetic term of  $\tilde{h}$ has the asymptotic form
\be \label{hTildeAsymptotic}
\mathcal{K}_{\tilde{h}} \overset{\xi_1\rightarrow \infty}{\rightarrow} \frac{13}{2 \tilde{h}^2} \;.
\ee
We are now in position to compare our findings with the action~\eqref{actionHiggsMetricEinstein} of single-field metric Higgs inflation. Upon identifying $h$ with $\tilde{h}$, we see from~\eq~\eqref{actionhTilde} that the potentials coincide and the coefficient of the kinetic terms only differ slightly in their numerical prefactors: the coefficient $6$ of metric Higgs inflation is replaced by $13/2$, as is evident from the asymptotic form~\eqref{hTildeAsymptotic}. Correspondingly, the model~\eqref{actionhTilde} yields the inflationary indices\,\footnote
{For example, this can be read off from~\cite{Shaposhnikov:2020gts} after identifying (in the notation and nomenclature of this article) $\x_\eta^2/\x^2 = 13/12$ in the Nieh-Yan case.}
\be \label{predictionsPrototype}
n_s \approx 1- \f 2 N \ ,~~~r \approx \f {13} {N^2} \ .
\ee
We observe that the spectral index is identical to the one of metric Higgs inflation and the tensor-to-scalar ratio is slightly larger.

We checked the above findings with a full numerical analysis of the inflationary dynamics, where we use $U/\epsilon = 5 \cdot 10^{-7}$ for the CMB normalization \cite{Akrami:2018odb} and choose $N=55$ as well as $\lambda=10^{-3}$ as typical values (see \cite{Shaposhnikov:2020gts}). For comparison, we first analyze the original model \eqref{actionHiggsMetric}, for which we obtain\,\footnote
{Formula \eqref{predictionsMetric}, which is derived to leading order in $1/N$, would give $r=0.00400$; more precise analytic results are available~\cite{Rubio:2018ogq}. The only goal of our present analysis, however, is to compare metric Higgs inflation as defined by \eq \eqref{actionHiggsMetric} with our model \eqref{actionhTilde}. This is possible as long as we employ the same approximations in both theories.}
\be
n_{s,\, \text{num}}^{\rm HI} = 0.965 \ ,~~~r_{\text{num}}^{\rm HI} = 0.00351 \ .
\ee
Repeating the same procedure in our model~\eqref{actionhTilde} (without performing any approximations in $\mathcal{K}_{\tilde{h}}$) yields
\be
n_{s,\, \text{num}} = 0.965 \ ,~~~r_{\text{num}} = 0.00379 \ .
\ee
We see that the spectral indices coincide and the ratio $r_{\text{num}}/r_{\text{num}}^{\rm HI}=1.080$, which is very close to $13/12$. This confirms the validity of formula \eqref{predictionsPrototype}.

To study the geometry of field-space, we work with $Z= M_P^2/(\x_1 \tilde{h}^2)$. Then we get from \eq \eqref{actionhTilde}
\be
\label{eq:action_Z}
S \approx\int\diff^4 x \sqrt{g}\left[ \f{M_P^2}{2} \mathring R - \f{M_P^2}{2}\l(\g_{\rho\rho} (\p_\m \tilde\rho)^2+\mathcal{K}_{Z} (\p_\m Z)^2\r)  -\f{\lambda M_P^4}{4 \x_1^2} \l(1-Z\r)^2 \right] \ ,
\ee
where $\mathcal{K}_{Z} = (\f{\diff \tilde{h}}{\diff Z})^2 \mathcal{K}_{\tilde{h}}$. Now we can compute the field-space curvature (see \eq \eqref{eq:curv_planck}) and the result is shown in~\fig~\ref{fig:curv_Z}. We conclude that $ M_P^2|\kappa| < 4$ for all relevant field values. In particular, the limiting cases are $ M_P^2\kappa = -2$ for $Z = 0$ as well as $M_P^2\kappa = 4$ for $Z=1$, corresponding to the limits of high energies and the electroweak epoch, respectively. In summary, we conclude that our model \eq \eqref{prototype} fulfills our conjecture. Conformality is preserved up to the Planck scale, i.e. $\Lambda_{\rm conf} \simeq M_P$, and in agreement with \eq \eqref{conjecture}, the predictions \eqref{predictionsPrototype} are close to their counterparts \eqref{predictionsMetric} of single-field metric Higgs inflation.

Finally, we establish a direct connection of field-space curvature and inflationary predictions. For $\x_1\gg 1$, our theory in the form \eqref{eq:action_Z} becomes 
\be
\label{eq:action_Z_approx}
S \approx\int\diff^4 x \sqrt{g}\left[ \f{M_P^2}{2} \mathring R - \f{M_P^2}{2}\l( 4\x_1^5 Z (\p_\m\tilde\rho)^2  +\f{13}{8} \f{(\p_\m Z)^2)}{Z^2}\r) -\f{\lambda M_P^4}{4 \x_1^2} \l(1-Z^2\r)^2 \right] \ ,
\ee
where we used that during inflation $\tilde{h}$ remains below $M_P$ and correspondingly $Z\gtrsim \mathcal O(1/\x_1)$. It is straightforward to compute the field-space curvature 
\be
M_P^2\kappa\approx -\f{4}{13} \ .
\ee
Now we can use \eq \eqref{rFromCurvature}, which is approximately valid for intervals of constant $\kappa$. Plugging in the value $-4/13$, we arrive at the tensor-to-scalar ratio $r=13/N^2$, in agreement with \eq \eqref{predictionsPrototype}. In summary, we conclude that both conformal properties and inflationary predictions can be deduced from the field-space curvature $\kappa$. In particular, an upper bound on $\kappa$ from the requirement of approximate conformal invariance leads to a lower bound on the tensor-to-scalar ratio.

Before turning to the conclusions, we shall evaluate the cutoff scale, above which perturbation theory breaks down. We can read it off from the potential as the energy scale that suppresses higher-dimensional operators. As is evident from \eq \eqref{actionhTilde}, we get
\be \label{cutoff}
	\Lambda_{\rm cutoff} \approx \frac{M_P}{\sqrt{\xi_1}} \;.
\ee
Taking into account effects of the non-canonical kinetic terms in \eq \eqref{actionhTilde} does not lower the cutoff scale since the field-space curvature is bounded as $M_P^2|\kappa| \lesssim 1$ (see discussion in \cite{Mikura:2021clt,Karananas:2022byw}). We remark that the cutoff scale can depend on the background value of fields \cite{Bezrukov:2010jz} and the result \eqref{cutoff} only refers to vacuum. Moreover, we have only considered a biscalar theory; including other fields, in particular longitudinal gauge bosons, may also influence the cutoff scale \cite{Bezrukov:2010jz,Shaposhnikov:2020geh,Mikura:2021clt,Karananas:2022byw}. We leave an investigation of these points for future work.

\section{Conclusions}
\label{sec:conclusions}

In this paper, we first discussed in detail the constraints that linearly as well as nonlinearly realized conformal invariance imposes on the dynamics of multiscalar field theories. We showed that the target manifolds are endowed with a rather specific geometry. To reiterate our finding,  there exists an appropriate set of variables in which the field-space metric of such conformal theories not only becomes block-diagonal (as is the case for theories invariant under dilatations only) but its uppermost left component---corresponding to the dilaton field ``coordinate''---is unity. In other words, there always exists an appropriate set of variables such that the dilaton has a canonical kinetic term and no kinetic mixings with the rest of the fields. From the phenomenological point of view, the interesting situation corresponds to biscalar theories for which conformal symmetry fixes the target manifold to actually be flat.

We then presented how the inclusion of Einstein-Cartan gravity may be effectuated in a manner that preserves invariance under global dilatations. Deviating from the commonly used metric incarnation of General Relativity, one has to and actually should account for invariants constructed out of torsion, too. This has to be done with care as shown in our previous works~\cite{Karananas:2021zkl,Karananas:2021gco} 
 (see also~\cite{Shaposhnikov:2020gts,Shaposhnikov:2020frq,Shaposhnikov:2020aen,Rigouzzo:2023sbb}), where we devised a comprehensive set of criteria for constructing actions that encompass EC gravity and matter fields and propagate only the two polarizations of the massless graviton in their (purely) gravitational sector. In general, the presence of (large) nonminimal coupling(s) translates into gravitational contributions finding their way into the kinetic sector. This breaks conformal invariance at energies (significantly) below the Planck mass (whereas scale symmetry is preserved). We showed how to remedy the situation by formulating a condition ensuring approximate conformality of the resulting theories up to the Planck scale, where the theory becomes scale invariant ``gravitationally.''
 
Our motivation for imposing this requirement is twofold. On the one hand, conformal invariance can improve the high-energy limit of the theory, by opening up a perspective for ``self-completion'' above the naive perturbative cutoff scale. On the other hand, subjecting a theory to such a condition reduces the built-in arbitrariness due to the numerous parameters that emerge in the EC formulation of GR. Investigating several concrete examples, we found indications that the situation may even be better than expected and -- along with similar behavior noticed before ~\cite{Karananas:2016kyt} -- this led us to conjecture that the requirement of approximate conformality up to the Planck scale implies nearly model-independent statements about inflationary observables, which turn out to be close to the predictions of single-field metric Higgs inflation. How far this universality goes and if it holds in all parts of parameter space remains to be determined.

\section*{Acknowledgements}  We are grateful to Andrey Shkerin and Inar Timiryasov for useful comments on the manuscript. The work of M.S. was supported in part by the Generalitat Valenciana grant PROMETEO/2021/083. S.Z. acknowledges support of the Fonds de la Recherche Scientifique - FNRS.

 \appendix

\section{Going beyond polynomial coefficient functions}
\label{app:deviation}

In this appendix we shall briefly discuss what would happen if we deviated from the requirement of having the various coefficient functions polynomial in the fields $\phi_1$ \& $\phi_2$. In other words, we will investigate what changes if the condition \textit{i)} of our conjecture is dropped.  To make things clear, we shall confine ourselves to an arguably extreme situation in which the curvature of the field-manifold is constant and equal to the Planck area for all field values.  It will become clear that the inflationary predictions stop being unique, even though the curvature has been fixed and conformality of the kinetic sector is preserved until the Planck scale. 

We start from (\cf \eq \eqref{eq:curv_planck})  
\be
M_P^2\kappa =  \f{\g_{\rho\rho}'\G'-2\G  \g_{\rho\rho}''}{2\G^2}  = 1 \ ,
\ee
with $\G= \g_{\t\t}\g_{\rho\rho}-\g_{\rho\t}^2$; assuming that $\g_{\rho\rho}\neq {\rm const.}$ we  obtain \cite{Karananas:2016kyt}
\be
\G= -\f{\g_{\rho\rho}'^2}{2\g_{\rho\rho}-c} \ ,
\ee
where $c$ is an arbitrary (dimensionless) constant. Therefore, the inflaton's kinetic function reads as (\cf \eq \eqref{Kprototype})
\be
\mathcal K = \f{\G}{\g_{\rho\rho}}= -\f{\g_{\rho\rho}'^2}{\g_{\rho\rho}(2\g_{\rho\rho}-c)} \ .
\ee
Even though the arbitrariness has been reduced, we notice that the behavior of $\mathcal K$ is not unique but depends on whether or not one can neglect $c$.  As long as inflation takes place for field values such that  $\g_{\rho\rho}> c$, then the canonical field $\chi$ follows from an exponential map, $\chi \sim \ln \g_{\rho\rho}$, and the predictions mimic the ones of metric Higgs inflation. If, on the other hand, $\g_{\rho\rho}< c$, then the inflationary dynamics is more intricate and depend on $c$, too. 

A particular choice of functions yielding this kind of behavior can for instance be the following ``parity preserving'' situation in \eq \eqref{eq:action_conformal_torsion}:
\be
G_{aa} = \z^a_\s = \z^a_\t = 0 \ .
\ee
Additionally, we take
\be
G_R = \x_1 \cos^2\t +\x_2 \sin^2\t \ ,~~~f_{\t\t} = \f 3 2 \l(\f{G_R'}{G_R}\r)^2 \f{f(\t)}{1+6G_R} \ , 
\ee
with $f(\t)$ to be determined. From this choice, we obtain 
\be
\g_{\rho\rho} = 6+\f{1}{G_R} \ ,~~~\g_{\rho\t} = 3 \f{G_R'}{G_R} \ ,~~~\g_{\t\t} =\f 3 2 \l(\f{G_R'}{G_R}\r)^2\l( 1+ \f{f(\t)}{1+6G_R}\r) \ ,
\ee
which translates into a rather involved $f(\t)$ 
\be
f(\t) = -1 - \f{2}{3(2+(c+12)G_R)} \ .
\ee
Clearly, the resulting $f_{\t\t}$ deviates from $f_{\t\t}=1$, which would be the only choice allowed according to our criteria \emph{i)} - \emph{iii)}.\footnote{We note that equal non-minimal couplings, $\xi_1=\xi_2$, are not viable in this example because then $G_R={\rm const.}$, so $f_{\t\t}=0$ and the kinetic term of $\theta$ would vanish.}

\bibliographystyle{utphys}
\bibliography{Refs}

\end{document}